\documentclass[twoside,11pt]{article}

\usepackage{blindtext}
\usepackage{jmlr2e}

\usepackage{color} 
\usepackage{amsmath}
\usepackage{amssymb,epsfig}

\include{pythonlisting}
\usepackage{enumitem}   
\usepackage{xr}

\usepackage{lastpage}
\jmlrheading{25}{2024}{1-\pageref{LastPage}}{10/22; Revised
6/24}{6/24}{22-1233}{Oliver Dukes, Stijn Vansteelandt and David Whitney}
\ShortHeadings{title}{Dukes, Vansteelandt and Whitney}

\ShortHeadings{Doubly Robust Inference for Double Machine Learning}{Dukes, Vansteelandt and Whitney}
\firstpageno{1}

\usepackage{xr}

\newcommand{\bea}{\begin{eqnarray}}
\newcommand{\eea}{\end{eqnarray}}

\newcommand\norm[1]{\left\lVert#1\right\rVert}

\newtheorem{ass}{Assumption}

\begin{document}

\title{On Doubly Robust Inference for Double Machine Learning in Semiparametric Regression}

\author{\name Oliver Dukes \email oliver.dukes@ugent.be \\
       \addr Department of Applied Mathematics, Computer Science and Statistics\\
       Ghent University\\
       9000 Ghent, Belgium
       \AND
       \name Stijn Vansteelandt \email stijn.vansteelandt@ugent.be \\
       \addr Department of Applied Mathematics, Computer Science and Statistics\\
       Ghent University\\
       9000 Ghent, Belgium
       \AND
       \name David Whitney \email david.e.whitney@gsk.com \\
       \addr GSK\\
       Gunnels Wood Road, Stevenage, SG1 2NY, U.K.}

\editor{Vanessa Didelez}

\maketitle

\begin{abstract}
Due to concerns about parametric model misspecification, there is interest in using machine learning to adjust for confounding when evaluating the causal effect of an exposure on an outcome. Unfortunately, exposure effect estimators that rely on machine learning predictions are generally subject to so-called plug-in bias, which can render naive $p$-values and confidence intervals invalid. Progress has been made via proposals like targeted minimum loss estimation and more recently double machine learning, which rely on learning the conditional mean of both the outcome and exposure. Valid inference can then be obtained so long as both predictions converge (sufficiently fast) to the truth. Focusing on partially linear regression models, we show that a specific implementation of the machine learning techniques can yield exposure effect estimators that have small bias even when one of the first-stage predictions does not converge to the truth. The resulting tests and confidence intervals are doubly robust.  We also show that the proposed estimators may fail to be regular when only one nuisance parameter is consistently estimated;  nevertheless, we observe in simulation studies that our proposal can lead to reduced bias and improved confidence interval coverage in moderate-to-large samples.
\end{abstract}


\section{Introduction}

In recent years, there has been a resurgence of interest in so-called doubly robust estimators \citep{robins_comments_2001}. These estimators require estimation of two separate nuisance parameters, and are consistent so long as one of the nuisance parameters is consistently estimated. When these estimators were introduced, focus was initially on postulating parametric working models, where at least one of the models should be correctly specified \citep{bang2005doubly}. However, it turns out that doubly robust methods are additionally advantageous when the nuisance parameters are estimated nonparametrically; for example, using modern machine learning techniques. If both are estimated consistently at sufficiently fast rates and sample splitting/cross-fitting is used, then standard inference can be performed on the target parameter based on a simple sandwich estimator or the non-parametric bootstrap. This is in spite of the nonparametric estimators having a potentially large bias and an unknown or intractable asymptotic distribution. Such ideas underpin developments in targeted minimum loss estimation (TMLE) \citep{van_der_laan_targeted_2011}, and double machine learning \citep{chernozhukov2018double}. 
Doubly robust estimators are a subclass of the methods based on orthogonal estimating functions that arise in semiparametric theory and which are reviewed in \citet{chernozhukov2018double}; an advantage of this particular subclass is that parametric-rate inference for the target parameter can be achieved even when one nuisance parameter is estimated  at a slow rate, so long as this is compensated by fast estimation of the other nuisance parameter. 

However, the possibility of parametric-rate inference for a target parameter after data-adaptive estimation of nuisance parameters is jeopardised when one of the nuisance parameters is not estimated consistently. A parametric-rate test has power to detect local alternatives that go to 0 like $n^{-1/2}$ and a parametric-rate confidence interval has width of the order $n^{-1/2}$. Although machine learning methods are more flexible than traditional parametric estimators, they may still be guilty of extrapolation in causal inference and missing data problems e.g. where treated and untreated patients are very different in terms of the distribution of their covariates. Although the doubly robust estimator remains consistent when just one of the nuisance parameters is not estimated consistently, its bias and rate of convergence will tend to be determined by the consistent nuisance parameter estimator. The behaviour of the latter, which is often poorly understood, propagates into the asymptotic distribution of the doubly robust estimator,  such that its asymptotic linearity (and thus asymptotic normality) no longer holds. In the context of estimating the expected counterfactual outcome under a given treatment, \citet{van2014targeted} describes how to amend estimation of the nuisance parameters within the TMLE procedure, in order to ensure that the resulting estimator of the mean is asymptotically linear if either a model for the probability of treatment or an outcome imputation model is correctly specified, even when nonparametric methods are used to estimate the nuisance parameters. 
As a consequence, doubly robust hypothesis tests and confidence intervals can be constructed. \citet{benkeser2017doubly} simplify this nuisance estimation procedure and show that standard `one-step' doubly robust estimators cannot be easily adapted to guarantee doubly robust inference, as opposed to TMLE. \citet{diaz2017doubly} and \citet{diaz2019statistical} extend this work respectively to randomised trials with missing outcome data and observational studies with survival outcomes. 

Several important questions remain. Firstly, can doubly robust inference be extended to parameters in more general semi/nonparametric models? Existing examples based on TMLE have focused on the expected counterfactual outcome (or closely related parameters). 
Secondly, the proposal of  \citet{benkeser2017doubly} involves nonparametric estimation of two additional nuisance parameters, but the necessary conditions on their estimators remain unclear. In particular, the interplay between their rates of convergence and those of the estimators (on which they depend) of the outcome and treatment models requires further exploration. Thirdly, the variance formula in  \citet{benkeser2017doubly} differs from the usual one of $n^{-1}$ times the sample variance of the efficient influence function. This raises the question of whether the proposed doubly robust estimators are non-regular; their asymptotic distribution may be highly sensitive to small changes in the data-generating law. Confirmation of this conjecture, along with guidance about when these methods would be expected to deliver poor performance, is therefore of interest.

In this work, we address these questions directly. We show how to adapt standard doubly robust estimators in order to yield doubly robust tests and confidence intervals in semiparametric regression problems, and spell out the implications for the `variance-weighted' treatment effect \citep{crump2006moving,robins2008higher}.  
In this context, a test is doubly robust if its type I error rate is asymptotically correct so long as at least one out of two nuisance parameters is consistently estimated; a confidence interval is doubly robust if it attains its advertised coverage (asymptotically) so long as at least one out of two nuisance parameters is consistently estimated. We also give a detailed theoretical treatment of kernel estimators of the additional nuisance parameters, exploring under what conditions they obtain the rates necessary to achieve doubly robust inference. Further, we describe the regularity properties of the novel doubly robust estimators. Unlike \citet{benkeser2017doubly}, we do not rely on Donsker conditions on the nuisance parameters estimators but rather use cross-fitting as in cross-validated TMLE \citep{van_der_laan_cross-validated_2011} and double machine learning \citep{chernozhukov2018double}, which we explicitly incorporate into our proofs. \citet{dukes2020doubly} and \citet {dukes2020inference} also obtain doubly robust inference for the partially linear model when (potentially) high-dimensional parametric models are postulated for the nuisance parameters; that approach is distinct from the one in this paper, given that it is reliant on this parametric structure of the working models as well as specific sparse estimators of the regression parameters. In comparison, the framework in the current paper is generic in that it can allow for arbitrary nonparametric estimators of the propensity score and conditional outcome mean. The work is also relevant to the growing literature on conditional independence testing; our score test can be viewed as an extension of the Generalised Covariance Measure (GCM) test proposed in \citet{shah2020hardness}.

\section{Review of doubly robust $Z$-estimators}
Let $W$ denote a random data unit with a distribution $P$ that is contained in a statistical model $\mathcal{M}$.  Consider an estimating function $\psi(W;\theta,\eta)$ for a finite dimensional target parameter $\theta_0$, such that $\theta_0$ is the unique solution to $\int \psi(w;\theta,\eta_0)dP(w)=0$. Here, $\eta_0$ is a nuisance parameter that in a non/semiparametric model may be infinite-dimensional. Since $\eta_0$ is generally unknown, it is usually substituted by an estimator $\hat{\eta}$; let $\eta^*$ refer to its probability limit. Suppose we have $n$ independent and identically distributed copies of $W$: $W_1,...,W_n$. Then a $Z$-estimator $\hat{\theta}$ of $\theta_0$ can be constructed as the solution to the equations
\[0=\frac{1}{n}\sum^n_{i=1}\psi(W_i;\theta,\hat{\eta}).\]
 In what follows, $\mathbb{P}_n$ denotes the empirical measure and we use the notation $Pf=\int f(x) dP(x)$ where $f$ is fixed. When applying the $\mathbb{P}_n$ and $P$ operators to $\psi(W;\theta,\eta)$, we will usually suppress dependence on $W$ e.g. $P\psi(\theta_0,\eta_0)=0$. We also define $V=-\partial P\psi(\theta,\eta^*)/\partial \theta |_{\theta=\theta_0}$ and we assume that $V$ is invertible. Supposing that $\hat{\eta}$ is obtained from an auxiliary sample, then one can show (see e.g. Theorem 5.31 of \citet{van2000asymptotic}) that 
\begin{align*}
\hat{\theta}-\theta_0=&V^{-1}(\mathbb{P}_n-P)\psi(\theta_0,\eta^*)+V^{-1}P\psi(\theta_0,\hat{\eta})+o_P(n^{-1/2}+\norm{P\psi(\theta_0,\hat{\eta})})
\end{align*}
where $\norm{\cdot}$ is the Euclidean norm; in places we will also use  $\norm{\cdot}_{P,q}=P(|f|^q)^{1/q}$. Therefore, the asymptotic behavior of  $\hat{\theta}$ depends on $\hat{\eta}$ via the so-called `drift' term $P\psi(\theta_0,\hat{\eta})$, which is the remainder from a linear expansion of $\hat{\theta}$.

Certain statistical problems enable the construction of `doubly robust' estimating functions. These depend on two  nuisance parameters $m_0$ and $g_0$ that comprise $\eta_0$, which are respectively estimated as $\hat{m}$ and $\hat{g}$; let $m^*$ and $g^*$ refer to the limits of these estimators. We will assume $m_0$ and $g_0$ are variation independent, in the sense that constraints on one of the two laws does not place restrictions on the other. Then $\int \psi(w;\theta_0,\eta^*)dP(w)=0$ if either $m_0=m^*$ or $g_0=g^*$. In this paper, we are interested in estimators whose drift can be written as $P\{d(\hat{g}-g_0)(\hat{m}-m_0)\}$ for some $d=d(W)$ that is upper bounded. The term $d$ typically arises via a linearization-type argument used to express the drift in terms of the errors in $\hat{g}$ and $\hat{m}$ (rather than transformations of these quantities); as we will see below, for certain estimating functions this may equal 1. By application of the Cauchy–Schwarz inequality, it follows that the drift can be upper bounded by a term proportional to
\[\norm{\hat{m}-m_0}_{P,2}\norm{\hat{g}-g_0}_{P,2}.\]
This implies that so long as both $\hat{m}$ and $\hat{g}$ converge to the truth (in the root-mean-squared error sense) at a rate faster than $n^{-1/4}$, then the drift is $o_P(n^{-1/2})$ and the resulting estimator $\hat{\theta}$ is asymptotically linear. In low-dimensional settings, classical nonparametric regression estimators may meet these rates, as may certain statistical/machine learning methods under structured assumptions on the underlying smoothness/sparsity of the estimated functional. 

The combination of doubly robust methods with flexible, data-adaptive estimation of nuisance parameters thus underpins recent developments in targeted learning \citep{van_der_laan_targeted_2011}, and double machine learning \citep{chernozhukov2018double}. This in turn builds on earlier foundational work on efficient estimation in non/semiparametric models \citep{bickel1993efficient,pfanzagl2012contributions,ibragimov2013statistical}. Indeed, doubly-robust $Z$-estimators form a subset of a general class of estimating equation-based approaches \citep{van2003unified}, involving estimating functions that are \textit{Neyman-orthogonal} with respect to $\eta_0$, using the terminology of \citet{chernozhukov2018double}. This means that the Gateaux derivative operator (taken with respect to the infinite-dimensional nuisance parameters) vanishes when the estimating function is evaluated at the true parameter values. We note that there exist other frameworks for constructing estimators for which an expansion yields a linear term and a second-order remainder (specifically, one-step estimation and TMLE). We adopt the estimating equations formulation as it will prove more natural in the context of our running example. A specific benefit of doubly robust estimators is that the cross-product structure of the remainder allows $\hat{m}$ to converge slowly (potentially even slower than $n^{-1/4}$), so long as $\hat{g}$ converges quickly. This property is known as `rate double-robustness' \citep{smucler_unifying_2019}.

\begin{example}{(Partially linear regression model).}
Consider the model 
\begin{align}\label{plm}
Y=\theta_0A+m_0(L)+\epsilon, \quad \mathbb{E}(\epsilon|A,L)=0
\end{align}
where $m_0(L)=\mathbb{E}(Y|A=0,L)$ and $A$ is one-dimensional. This model is defined by the key restriction that $A$ acts linearly on the conditional mean of $Y$, and this effect is not modified by $L$. If the identification conditions for causal inference in observational studies hold (in particular, that there is no unmeasured confounding) and $A$ is dichotomous, then $\theta_0$ can be interpreted as the average causal effect of $A$ on $Y$ a.k.a. $\mathbb{E}(Y^1-Y^0)$, where $Y^a$ denotes the counterfactual outcome that would be observed were $A$ set to level $a$.

A doubly robust estimator of $\theta_0$ can be constructed using the estimating function:
\[\psi(W;\theta_0,\eta_0)=\{A-g_0(L)\}\{Y-\theta_0A-m_0(L)\}\]
where $g_0(L)=\mathbb{E}(A|L)$. In this case, one can estimate $\theta_0$ in closed-form as 
\begin{align}\label{dr_est}
\hat{\theta}=\frac{\sum^n_{i=1}\{A_i-\hat{g}(L_i)\}\{Y_i-\hat{m}(L_i)\}}{\sum^n_{i=1}\{A_i-\hat{g}(L_i)\}A_i}
\end{align}
\citep{robins_estimating_1992}. By the previous expansion, it follows immediately that the drift term for $\hat{\theta}$ is asymptotically negligible so long as $\norm{\hat{m}-m_0}_{P,2}\norm{\hat{g}-g_0}_{P,2}=o_P(n^{-1/2})$. Compare this with an alternative, `singly robust' estimator 
\begin{align}\label{g-estimator}
\tilde{\theta}=\frac{\sum^n_{i=1}\{A_i-\hat{g}(L_i)\}Y_i}{\sum^n_{i=1}\{A_i-\hat{g}(L_i)\}A_i}
\end{align}
\citep{robins_estimating_1992}. The drift for $\tilde{\theta}$ is asymptotically negligible if $\norm{\hat{g}-g_0}_{P,2}=o_P(n^{-1/2})$. This will not be the case when nonparametric/data-adaptive estimators of $\hat{g}(L)$ are used; hence the drift for $\hat{\theta}$ may be asymptotically negligible whilst the drift for $\tilde{\theta}$ will often fail to be.
\end{example}

We have so far overlooked the `empirical process' terms that arise in expansions for $Z$-estimators. In order for $\hat{\theta}$ to be asymptotically linear,  we require that  
\begin{align*}
&(\mathbb{P}_n-P)\psi(\hat{\theta},\hat{\eta})=(\mathbb{P}_n-P)\psi(\theta_0,\eta^*)+o_P(n^{-1/2}),
\end{align*} hence we must control the difference between the linear term and its empirical analogue. This is often done via Donsker conditions, which restrict the complexity of the nuisances and their estimators. An alternative is to use sample splitting. Suppose that $\theta_0$ is estimated from a separate part of the sample than that used to estimate $\eta_0$; it follows loosely e.g. from the proof of Theorem 25.57 in \citet{van2000asymptotic} that so long as $P\norm{\psi(\theta_0,\hat{\eta})-\psi(\theta_0,\eta^*)}^2=o_P(1)$, by Chebyshev's inequality, the empirical process term is $o_P(n^{-1/2})$. Given that it is unclear whether Donsker conditions hold for many common machine learning methods, sample splitting is appealing in its generality and transparency. However, this comes at the cost of loss of sample size, which has driven the development of cross-validated TMLE and cross-fitting \citep{van_der_laan_cross-validated_2011,chernozhukov2018double}. For example, a separate estimate of $\theta_0$ can be obtained in each split (using estimates of $g_0$ and $m_0$ from the other parts of the data) and then these can be averaged; asymptotically there should be no efficiency loss. In the above example, if the errors $Y-\theta_0-m_0(L)$ are homoscedastic, the drift is $o_P(n^{-1/2})$, and either Donsker conditions hold or cross-fitting is used, then the doubly robust estimator attains the semiparametric efficiency bound under the partially linear model.

\section{Inference when only one nuisance parameter is estimated consistently}

\subsection{Doubly robust consistency and asymptotic linearity}

Let us return to the drift term of the doubly robust estimator $\hat{\theta}$. Suppose that $\hat{g}$ converges to $g_0$, but $\hat{m}$ converges to some limit $m^*\neq m_0$. Then 
\begin{align*}
&P\psi(\theta_0,\hat{\eta})
\leq \norm{\hat{g}-g_0}_{P,2}O_P(1)=o_P(1)
\end{align*}
One can also show that the drift is $o_P(1)$ if $\hat{m}$ converges to $m_0$ but $g^*\neq g_0$. Hence the drift shrinks to zero if one of the two estimators converges to the truth, regardless of which one. Furthermore, the empirical process term will not be affected, and so long as one nuisance parameter is estimated correctly, the linear term continues to be mean zero and asymptotically normal. This property is known as \textit{doubly robust consistency}, since the consequence is that $\hat{\theta}$ is consistent so long as one of the nuisance estimators is also consistent. 

A stronger condition than doubly robust consistency is \textit{doubly robust asymptotic linearity} \citep{benkeser2017doubly}. This is necessary in order to obtain valid parametric-rate tests and confidence intervals, in addition to a consistent estimator of $\theta_0$. In this context, an estimator is doubly robust asymptotically linear if either $\norm{\hat{g}-g_0}_{P,2}=o_P(1)$ or $\norm{\hat{m}-m_0}_{P,2}=o_P(1)$ (or both) and the drift and empirical process terms are $o_P(n^{-1/2})$. The standard doubly robust estimator will not typically possess this property; assuming only e.g. that $\norm{\hat{g}-g_0}_{P,2}=o_P(n^{-\kappa})$ for $\kappa<1/2$, then the drift is also $o_P(n^{-\kappa})$. 
Without additional parametric structure, the drift is not typically asymptotically linear. Since the drift now contributes to the first order behaviour of the estimator, the resulting estimator of $\theta_0$ may inherit non-negligible bias from the data-adaptive estimator $\hat{m}$. 

\subsection{Expansion of the drift term}

Suppose that either $\hat{g}$ does not converge to $g_0$ or $\hat{m}$ does not converge to $m_0$. Then if the drift can be written as  $P\{d(\hat{g}-g_0)(\hat{m}-m_0)\}$, then one can further decompose it as
\begin{align*}
P\psi(\theta_0,\hat{\eta})&=P\{d(\hat{g}-g^*)(\hat{m}-m^*)\}+P\{d(\hat{g}-g^*)(m^*-m_0)\}\\&\quad+P\{d(g^*-g_0)(\hat{m}-m^*)\}\\
&=R_1+R_2+R_3
\end{align*}
So long as both estimators converge sufficiently fast to a limit, then $R_1=o_P(n^{-1/2})$. The key terms that characterise how the drift contributes to the first order behaviour of the estimator are thus $R_{2}$ and $R_{3}$; note that so long as one nuisance is consistently estimated, then one of these terms will equal zero. In the case of inference on the average counterfactual mean under treatment $A=a$, $\mathbb{E}(Y^a)=\mathbb{E}\{\mathbb{E}(Y|A=a,L)\}$ (or equivalently, a population mean with outcomes missing-at-random), \citet{van2014targeted} and \citet{benkeser2017doubly} show that when $m^*\neq m_0$, $R_2$ can be decomposed as:
\begin{align*}
R_2=\mathbb{P}_n\psi_2(\theta_0,\hat{\eta},\hat{\tau})
-(\mathbb{P}_n-P)\psi_2(\theta_0,\eta^*,\tau^*)
-P\psi_2(\theta_0,\hat{\eta},\hat{\tau})+o_P(n^{-1/2})
\end{align*}
where $\psi_2(\theta_0,\eta_0,\tau_0)$ is a mean-zero random variable,
$\tau_0$ is an additional nuisance parameter to be estimated and $\hat{\tau}$ is its estimator (assumed again to be constructed on a separate sample), with limiting value $\tau^*$. Note that the first term on the right hand side of the above equality can be estimated using the data and the second is a sample average of mean-zero random variables. Further, the authors show that $|P\psi_2(\theta_0,\hat{\eta},\hat{\tau})|$ can be upper bounded by a term proportional to $\norm{\hat{g}-g_0}_{P,2}\norm{\hat{\tau}-\tau_0}_{P,2}$, and hence is second-order so long as both $\hat{g}$ and $\hat{\tau}$ converge sufficiently quickly. A similar decomposition of $R_3$ in terms of a mean-zero random variable $\psi_3(\theta_0,\eta_0,\tau_0)$ is also available. 

The form of the additional nuisance parameter $\tau_0$ appears to be specific to the problem setting. Indeed, in the case of the parameter $\mathbb{E}\{\mathbb{E}(Y|A=a,L)\}$, multiple parametrisations (with potentially quite different properties) are available \citep{van2014targeted,benkeser2017doubly}. To give further insight, we show below the above strategy for decomposing the drift term also applies to the partially linear model.

\begin{example}{Partially linear regression model (continued)}\\
Suppose that $g^*(L)=g_0(L)$ but $m^*(L)\neq m_0(L)$, then $R_3=0$ and $R_2
=P\{\bar{G}(A-\hat{g})\}$ where $\bar{G}(L)=\mathbb{E}\{Y-\theta_0A-m^*(L)|g^*(L),\hat{g}(L)\}$. Then by adding and subtracting terms (and with some abuse of notation), we have,
\begin{align*}
R_2&=\frac{1}{n}\sum^n_{i=1}\hat{G}(L_i)\{A_i-\hat{g}(L_i)\}\\&\quad-(\mathbb{P}_n-P)\{G^*(A-g_0)\}-(\mathbb{P}_n-P)\{\hat{G}(A-\hat{g})-G^*(A-g_0)\}\\
&\quad+P\{(\bar{G}-G^*)(g_0-\hat{g})\}+P\{(G^*-\hat{G})(g_0-\hat{g})\}.
\end{align*}
where $G^*(L)=\mathbb{E}\{Y-\theta_0A-m^*(L)|g^*(L)\}$ and $\hat{G}(L)$ is an estimator of $G^*(L)$ (a specific approach is proposed in the following section). If we assume for now that the empirical process term and $P\{(\bar{G}-\hat{G})(g_0-\hat{g})\}$ are $o_P(n^{-1/2})$ then it follows that an estimator that solves the augmented equations
\[0=\frac{1}{n}\sum^n_{i=1}\{A_i-\hat{g}(L_i)\}\{Y_i-\theta A_i-\hat{m}(L_i)\}-\hat{G}(L_i)\{A_i-\hat{g}(L_i)\}\]
for $\theta$ would have a drift that can be upper bounded by 
$\norm{\hat{G}-G^*}_{P,2}\norm{\hat{g}-g_0}_{P,2}$. This is because after subtracting the empirical bias term $n^{-1}\sum^n_{i=1}\hat{G}(L_i)\{A_i-\hat{g}(L_i)\}$ from the original estimating equations for $\theta_0$, what remains from the above decomposition of $R_2$ is a mean zero, linear term plus $P\{(G^*-\hat{G})(g_0-\hat{g})\}$ and  an $o_P(n^{-1/2})$ remainder.

One can repeat these arguments under the scenario that $m^*(L)= m_0(L)$, $g^*(L)
\neq g_0(L)$, where now $R_3=P\{\bar{M}(Y-\theta_0A-\hat{m})\}$
and $\bar{M}(L)=\mathbb{E}\{A-g^*(L)|m^*(L),\hat{m}(L)\}$. If $M^*(L)=\mathbb{E}\{A-g^*(L)|m^*(L)\}$ and $\hat{M}(L)$ denotes its estimator, then applying the same reasoning as above, solving the equations
\[0=\frac{1}{n}\sum^n_{i=1}\{A_i-\hat{g}(L_i)\}\{Y_i-\theta A_i-\hat{m}(L_i)\}-\hat{M}(L_i)\{Y_i-\theta A_i-\hat{m}(L_i)\}\]
for $\theta$ would yield a drift that can be bounded by $\norm{\hat{M}-M^*}_{P,2}\norm{\hat{m}-m_0}_{P,2}$.
\end{example}
In the following sections, we will make specific proposals for estimating $\tau^*=\{G^*(L),M^*(L)\}$; note how these additional parameters will generally have reduced dimension compared with $g_0(L)$ and $m_0(L)$. Interestingly, the additional nuisance parameters that arise in the case of the expected counterfactual mean are either higher in dimension \citep{van2014targeted}, or higher in number; the proposal of \citet{benkeser2017doubly} requires estimation of three (as opposed to two) additional nuisance parameters. Note that the above decompositions of $R_2$ and $R_3$ apply for a broad class of semiparametric models where doubly robust estimators exist; it is trivial to show that it also applies to the partially log-linear model \citep{robins_comments_2001} as well as to partially linear and log-linear models with instrumental variables \citep{okui2012doubly}. However, as the following example shows, it does not apply to the important case of the partially logistic model \citep{tchetgen2010doubly}.

\begin{example}{Partially logistic model.}
Under a `no unmeasured confounding' assumption, the parameter $\theta_0$ indexing the partially logistic model \[logit\{Pr(Y=1|A,L)\}=\theta_0A+m_0(L)\]
encodes the \textit{conditional} (rather than marginal) causal log-odds ratio $\theta_0=logit\{Pr(Y^1=1|L)\}-logit\{Pr(Y^0=1|L)\}$, due to the non-collapsible link function. Let $\nu_0(L)=\mathbb{E}(A|Y=0,L)$ and $\nu^*(L)$ denote the limit of its estimator $\hat{\nu}(L)$; then the score function
\[\{A-\nu^*(L)\}\{Ye^{-\theta_0A-m^*(L)}-(1-Y)\}\]
obtained by \citet{tan2019doubly} has mean zero if either $m_0(L)=m^*(L)$
or $\nu_0(L)=\nu^*(L)$. With some abuse of notation, it follows from \citet{tan2019doubly} that the drift term under the partially logistic model is equal to
\[P\left\{(1-Y)\left(e^{m_0-\hat{m}}-1\right)(\nu_0-\hat{\nu})\right\}.\]
In the Appendix, we show that this can be rearranged to equal 
\begin{align*}
&P\left[\mathbb{E}\{Ye^{-\theta_0A-m^*}-(1-Y)|\nu_0,\hat{\nu}\}\frac{(A-\hat{\nu})(1-Y)}{\{1-\mathbb{E}(Y|\nu_0,\hat{\nu})\}}\right]\\
&+P\left[\mathbb{E}\{A-\nu^*|Y=0,m_0,\hat{m}\}\{Ye^{-\theta_0A-\hat{m}}-(1-Y)\}\right]
\end{align*}
plus a term that can be shown to be $o_P(n^{-1/2})$ if either $\nu_0(L)=\nu^*(L)$ or $m_0(L)=m^*(L)$ and each estimator converges sufficiently fast to a limit. Hence an expansion of the drift term under this model yields three (as opposed to two) additional nuisance parameters: $\mathbb{E}\{Ye^{-\theta_0A-m^*(L)}-(1-Y)|\nu_0(L),\hat{\nu}(L)\}$, $P\{Y=0|\nu_0(L),\hat{\nu}(L)\}$ and $\mathbb{E}\{A-\nu^*(L)|Y=0,m_0(L),\hat{m}(L)\}$, although one could directly model \[\frac{\mathbb{E}\{Ye^{-\theta_0A-m^*(L)}|\nu_0(L)\}}{1-\mathbb{E}
\{Y|\nu_0(L)\}}-1.\]
\end{example}

To summarise thus far, when one of the nuisances is estimated inconsistently, then the drift term may now contribute to the first order behaviour of $\hat{\theta}$. In decomposing the drift, additional bias terms arise which can be estimated with the data at hand. For example, in the case of the partially linear model, one might be tempted to construct the one-step augmented estimator
\begin{align}\label{1s_error}
\hat{\theta}-\frac{\sum^n_{i=1}\hat{G}(L_i)\{A_i-\hat{g}(L_i)\}}{\sum^n_{i=1}\{A-\hat{g}(L_i)\}A_i}-\frac{\sum^n_{i=1}\hat{M}(L_i)\{Y_i-\hat{\theta}A_i-\hat{m}(L_i)\}}{\sum^n_{i=1}\{A_i-\hat{g}(L_i)\}A_i}
\end{align}
where one would expect the bias terms $n^{-1}\sum^n_{i=1}\hat{G}(L_i)\{A_i-\hat{g}(L_i)\}$ and $n^{-1}\sum^n_{i=1}\hat{M}(L_i)\{Y_i-\hat{\theta}A_i-\hat{m}(L_i)\}$ to serve a similar role in de-biasing the estimator $\hat{\theta}$ as did subtracting $n^{-1}\sum^n_{i=1}\{A_i-\hat{g}(L_i)\}\hat{m}(L_i)$ (after scaling) from the na\"ive estimator $\tilde{\theta}$. This requires estimation of the additional nuisances $G^*(L_i)$ and $M^*(L_i)$; \citet{benkeser2017doubly} note that these are univariate regression problems and thus fast rates may be available using nonparametric estimators (e.g. kernel methods with bandwidth parameter selected using cross validation). Nevertheless, as we will explore in Section \ref{kernel_sec}, they are also based on data-adaptive estimators $\hat{g}(L)$ and $\hat{m}(L)$, which has implications for the rates.

Unfortunately, as explained by \citet{benkeser2017doubly}, a proposal based on equation (\ref{1s_error}) would not generally lead to doubly robust asymptotic linearity, unless one knows \textit{a priori} whether $g_0(L)$ or $m_0(L)$ is consistently estimated. To illustrate why, consider again the setting where $g^*(L)=g_0(L)$ so that $M^*(L)=0$ but $m^*(L)\neq m_0(L)$. Since $R_2=0$, including a term proportional to $n^{-1}\sum^n_{i=1}\hat{M}(L_i)\{Y_i-\hat{\theta}A_i-\hat{m}(L_i)\}$ in the estimator will over-correct for the bias. 
This is because the first-order behaviour of $n^{-1}\sum^n_{i=1}\hat{M}(L_i)(Y_i-\hat{\theta}A_i-\hat{m}(L_i)\}$ will be determined by
\begin{align*}
P\{(M^*-\hat{M})(m_0-\hat{m})\}\leq \norm{\hat{M}-M^*}_{P,2}O_P(1)
\end{align*}
But as $M^*(L)$ is estimated non-parametrically, the rate of $\hat{M}(L_i)$ may be  $n^{-1/2}$ or likely slower, and so the right hand side of the above inequality fails to be $o_P(n^{-1/2})$. Although the misplaced correction term will converge to zero, it does not do this fast enough to retain asymptotic linearity. In the following section, we will remedy this.

\subsection{Proposal for score tests and confidence intervals in semiparametric regression}\label{proposal}

Before proposing how to obtain an estimator and confidence interval for $\theta_0$ in a partially linear model, as a first step we will begin by constructing a score test of the null hypothesis that $\theta=\theta_0$. Such a test could be obtained via the expected conditional covariance statistic
\begin{align}\label{unscaled}
\frac{1}{\sqrt{n}}\sum^n_{i=1}\{A_i-\hat{g}(L_i)\}\{Y_i-\theta_0A_i-\hat{m}(L_i)\}
\end{align}
after scaling by the empirical standard deviation of the estimated score $\{A_i-\hat{g}(L_i)\}\{Y_i-\theta_0A_i-\hat{m}(L_i)\}$. Note that the unscaled statistic will share the same drift as the estimator $\hat{\theta}$.  Following the arguments of the previous subsections, under the null hypothesis this statistic would have expectation zero so long as one of the nuisances is consistently estimated, but the test itself is not doubly robust. In what follows, we amend the statistic so that we have a test that is both doubly robust consistent and asymptotically linear. Specifically, we will propose a nuisance parameter estimation procedure that sets the bias terms in the previous expansion of the drift to zero, leaving a remainder that is second order. This procedure can be straightforwardly adjusted for parameters in the partially log-linear model, by replacing $Y-\theta_0 A$ with $Y\exp(-\theta_0A)$, or in semiparametric instrumental variable models, where the residual $A-\hat{g}(L)$ is replaced by the instrument minus its estimated conditional expectation.

We propose to estimate $M^*(L)$ and $G^*(L)$ non-parametrically using kernel smoothing. Specifically, if 
\[\varphi_j(L;x,h,\eta)=K\left(\frac{x-m^*(L)}{h}\right)\{A-g^*(L)\}^{j-1},\quad \quad j=1,2.\]
where $K$ is a kernel function and $h>0$ is a bandwidth parameter, then let us define the estimator
\begin{align*}
\hat{M}(x)
&=\hat{f}^{-1}_{\hat{m},n}(x)n^{-1}\sum^n_{i=1}\varphi_2(L_i;x,h,\hat{\eta}).
\end{align*}
Here, $x$ is a point in the interior of the support $\mathcal{X}$ of $m^*(L)$, $\eta=\{g^*(L),m^*(L)\}$ and 
\[\hat{f}_{\hat{m},n}(x)=n^{-1}\sum^n_{i=1}\varphi_1(L_i;x,h,\hat{\eta}).\]
The above is a version of the standard Nadaraya-Watson estimator, where both the regressor $m^*(L)$ and the dependent variable $A-g^*(L)$ must first be estimated from the data in an initial step. Redefining $x$ as a point in the interior of $\mathcal{X}$ of $g^*(L)$, then if 
\[\rho_j(L;x,h,\eta)=K\left(\frac{x-g^*(L)}{h}\right)\{Y-\theta_0A-m^*(L)\}^{j-1},\quad \quad j=1,2\] and $\hat{f}_{\hat{g},n}(x)=n^{-1}\sum^n_{i=1}\rho_1(L_i;x,h,\hat{\eta})$
then we can similarly define\\ $\hat{G}(x)=\hat{f}^{-1}_{\hat{g},n}(x)n^{-1}\sum^n_{i=1}\rho_2(L_i;x,h,\hat{\eta})$. We focus on kernel smoothers in part because of their optimality in certain univariate nonparametric regression contexts \citep{horowitz2009semiparametric}, and also because it is feasible to analyse them when outcome and regressors are themselves functions estimated nonparametrically \citep{mammen2012nonparametric,delaigle2016nonparametric}.

After obtaining the estimates $\hat{M}(L)$ and $\hat{G}(L)$, then the algorithm proposed below will update the predictions $\hat{g}(L)$ and $\hat{m}(L)$ as
\begin{align*}
&\hat{g}(L)+\hat{\alpha}\hat{G}(L)\quad\textrm{and}\quad\hat{m}(L)+\hat{\beta}\hat{M}(L)
\end{align*}
where $\hat{\alpha}$ and $\hat{\beta}$ are estimated respectively as the solutions to the equations:
\begin{align}
0&=\frac{1}{n}\sum^n_{i=1}\hat{G}(L_i)\{A_i-\hat{g}(L_i)-\alpha \hat{G}(L_i)\}\label{EE_G}\\
0&=\frac{1}{n}\sum^n_{i=1}\hat{M}(L_i)\{Y_i-\theta_0 A_i-\hat{m}(L_i)-\beta \hat{M}(L_i)\}.\label{EE_M}
\end{align}
Let $\alpha^*$ and $\beta^*$ be the limits in probability of $\hat{\alpha}$ and $\hat{\beta}$. By virtue of how $\hat{\alpha}$ and $\hat{\beta}$ are constructed, we set the bias terms $n^{-1}\sum^n_{i=1}\hat{G}(L_i)\{A_i-\hat{g}(L_i)-\hat{\alpha}\hat{G}(L_i)\}$ and $n^{-1}\sum^n_{i=1}\hat{M}(L_i)\{Y_i-\theta_0A_i-\hat{m}(L_i)-\hat{\beta}\hat{M}(L_i)\}$ (see Example 2) that may arise in the drift of the test statistic to zero, in a similar way to \citet{benkeser2017doubly}. 

We note that if $g_0(L)=g^*(L)$, then $\alpha^*=0$ and hence $\hat{G}(L)$ will not contribute in large samples to the test statistic (likewise, if $m_0(L)=m^*(L)$ then $\beta=0$). However, a concern is that if $m_0(L)=m^*(L)$ then the denominator of $\hat{\alpha}$ converges to zero since $\sum^{n_k}_{i=1}\hat{G}^2_k(L_i)$ converges to zero. We make two remarks on this point; firstly, note that $\hat{\alpha}$ is always multiplied by $\hat{G}(L)$, which also converges to zero in this scenario. This suggests that the potential divergence is not an issue so long as the sample average of $\hat{\alpha}\hat{G}(L)$ converges to zero. Results in Section 1.2 of the Appendix suggest this is indeed expected to be the case, and moreover that the behaviour of this sample average will be determined by the rate of $\hat{g}(L)$.  Secondly, we could alternatively define a bivariate fluctuation of $\hat{g}(L)$ (or $\hat{m}(L)$) that also includes an intercept, which means that even if $\hat{G}(L)$ is converging toward zero in probability the design matrix for the least squares update will have rank 1 in the limit. However the analysis of fitted values would follow along similar lines and we have not seen evidence of this consideration impacting inferences in our simulation studies.

Due to the concerns that Donsker conditions may not apply to the data-adaptive estimators of the nuisance parameters, we propose to use sample splitting combined with cross-fitting in constructing our tests. We describe how this can be done below:

\begin{enumerate}
\item Divide the sample into disjoint parts $I_k$ each of size $n_k=n/K$, where $K$ is a fixed integer (and assuming $n$ is a multiple of $K$). For each $I_k$, let $I^c_k$ denote all indices that are not in $I_k$. 
\item Obtain the machine learning estimates $\hat{g}^c_k(L)$ and $\hat{m}^c_k(L)$ from $I^c_k$.
\item Obtain the estimates $\hat{G}_k(L)$ and $\hat{M}_k(L)$ from $I_k$.
\item Obtain the estimates $\hat{\alpha}_k$ and $\hat{\beta}_k$ via solving the equations:
\begin{align*}
0&=\frac{1}{n_k}\sum_{i\in I_k}\hat{G}_k(L_i)\{A_i-\hat{g}^c_k(L_i)-\alpha \hat{G}_k(L_i)\}\\
0&=\frac{1}{n_k}\sum_{i\in I_k}\hat{M}_k(L_i)\{Y_i-\theta_0 A_i-\hat{m}(L_i;I^c_k)-\beta \hat{M}(L_i;I_k)\}].
\end{align*}
for $\alpha$ and $\beta$ from $I_k$.  
\item For all $i$ in $I_k$, obtain the score
\begin{align*}
\psi^*(W_i;\theta_0,\hat{\eta}^c_k,\hat{\tau}_k)=&\{A_i-\hat{g}^c_k(L_i)-\hat{\alpha}_k \hat{G}_k(L_i)\}\{Y_i-\theta_0 A_i-\hat{m}^c_k(L_i)-\hat{\beta}_k \hat{M}_k(L_i)\}\\
&-\hat{G}_k(L_i)\{A_i-\hat{g}^c_k(L_i)-\hat{\alpha}_k \hat{G}_k(L_i)\}\\
&-\hat{M}_k(L_i)\{Y_i-\theta_0 A_i-\hat{m}^c_k(L_i)-\hat{\beta}_k \hat{M}_k(L_i)\}
\end{align*}
and its average $\tilde{U}_{n_k}(I_k, I^c_k)=n_k^{-1}\sum_{i\in I_k}\psi^*(W_i;\theta_0,\hat{\eta}^c_k,\hat{\tau}_k)$.
\item Construct a test statistic of the null hypothesis that $\theta=\theta_0$ as 
\[\frac{\frac{1}{K}\sum^K_{k=1} \tilde{U}_{n_k}( I_k, I^c_k)}{\hat{\sigma}^2/\sqrt{n}}\]
where 
\begin{align*}
\hat{\sigma}^2=\frac{1}{K}\sum^K_{k=1}\left[ \frac{1}{n_k}\sum_{i\in I_k}\psi^*(W_i;\theta_0,\hat{\eta}^c_k,\hat{\tau}_k)^2\right]-\left\{\frac{1}{K}\sum^K_{k=1}\tilde{U}_{n_k}( I_k, I^c_k)\right\}^2.
\end{align*}
\end{enumerate}

Given that we can construct a valid test of the null hypothesis, we can invert the test in order to construct an estimator of $\theta_0$ along with a valid $100(1-\alpha)$\% confidence set. Specifically, one can conduct a grid search over the possible values of $\theta_0$; the point estimate will be the value for which the test statistic equals zero (and the $p$-value equals 1). Likewise, the limits of the interval can be obtained as the values of $\theta_0$ for which the scaled test statistic equals the $\alpha/2$ and $1-(\alpha/2)$ quantiles of the standard normal distribution. This approach is computationally intensive as it requires a separate estimate of the $m_0(L)$ for each choice of $\theta_0$. Nevertheless, in related work on doubly robust inference, score tests and inverted score test-based confidence intervals have been seen to have theoretical advantages (in terms of weaker assumptions on nuisance parameter estimators) and good empirical performance relative to Wald-based tests/intervals \citep{dukes2020doubly,dukes2020inference}.

\subsection{Connections with conditional independence testing}

Our work is relevant to the problem of testing for independence between two variables $A$ and $Y$ given a random vector $Z$. The literature on conditional independence testing has grown in the previous years, due to its relevance in causal discovery and high-dimensional statistics. With $\hat{G}(L_i)$ and $\hat{M}(L_i)$ set to zero, our proposed score test would reduce to the GCM test in \citet{shah2020hardness}; see also the $g$-null test of \citet{robins1986new}. Although \citet{shah2020hardness} show that no non-trivial conditional independence test exists that also possesses valid size under the conditional independence null, their approach is seen to control type I error and have reasonable power over range of data-generating processes (both theoretically and empirically). Our work can hence be viewed as a doubly robust extension of the GCM test. An additional difference between the proposals is that we use sample splitting and cross-fitting for the nuisance parameters; this controls remainder terms in our proofs when either $m_0$ or $g_0$ is estimated inconsistently.

\section{Theoretical results}

\subsection{Doubly robust asymptotic linearity of the test statistic}

We will now describe the theoretical properties of the proposed score test statistic from the previous section. Although formal results are developed for testing, they can be extended for estimators and confidence intervals as these can be constructed via the inversion of the test. For ease of exposition, the results are developed for the partially linear model, but can be straightforwardly extended to the other semiparametric regression models discussed above, and trivially hold for the `expected conditional covariance' parameter discussed e.g. in \citet{robins2008higher}. This is because in all of these examples, doubly robust estimators are known to exist and the resulting drift term decomposes in the same way (unlike e.g. for the partially logistic model). Before giving the main theorem, we list the key assumptions.

Introducing some further notation, let $f$ be a function belonging to a particular class $\mathcal{F}$; then 
$\mathbb{G}_n(f)=\sqrt{n}(\mathbb{P}_n-P)f$ and we will use $\mathbb{P}_{n,k}$ and $\mathbb{G}_{n,k}$ to refer to the sample average and empirical process respectively in the $k$-th split of the data. In order to make theoretical guarantees on our estimators, then we will need to place certain restrictions on the complexity of the (infinite-dimensional) nuisance parameters. We will use $F$ to denote an envelope function for $\mathcal{F}$; this is a function where $F(x)\geq|f(x)|$ for every $f \in \mathcal{F}$ and $x$ in the support of $X$. The uniform-entropy integral is
\[J(\delta,\mathcal{F},L_2)=
\int^\delta_0\sup_Q\sqrt{1+\log N(\epsilon \norm{F}_{Q,2},\mathcal{F},L_2(Q))}d\epsilon\]
where we take the supremum over all finitely discrete measures $Q$, and $L_2(Q)$ is the $L_2$ space under a probability measure $Q$ with semi-metric $\norm{f}_{Q,2}$ for any $f \in \mathcal{F}$. Also, $N(\epsilon \norm{F}_{Q,2},\mathcal{F},L_2(Q))$ is the $L_2(Q)$ covering number a.k.a. the minimum number of balls of radius $\epsilon$ in $L_2$ space required to cover the function class $\mathcal{F}$ with the envelope $F$ (we use the distance $\norm{\cdot}_{Q,2}$), and hence represents the complexity of the function class. Also, we redefine $\bar{M}(L)=\mathbb{E}\{A-g^*(L)|m_0(L),\hat{m}^c_k(L),\hat{\beta}_k\hat{M}_k(l)\}$ and $\bar{G}(L)=\mathbb{E}\{Y-\theta_0A-m^*(L)|g_0(L),\hat{g}^c_k(L),\hat{\alpha}_k\hat{G}_k(L)\}$.

We will denote the relevant score functions as:
\begin{align*}
\psi_1(W;\theta_0,\eta,\tau)&=\{A-g^*(L)-\alpha G^*(L)\}\{Y-\theta_0 A-m^*(L)-\beta M^*(L)\}\\
\psi_2(W;\theta_0,\eta,\tau)&=M^*(L)\{Y-\theta_0 A-m^*(L)-\beta M^*(L)\}\\
\psi_3(W;\theta_0,\eta,\tau)&=G^*(L)\{A-g^*(L)-\alpha G^*(L)\}.
\end{align*}
We are now in a position to list the main assumptions:

\begin{ass}\label{con_ML}{Convergence of nuisance parameter estimators to the truth}
\item 
For every $L$, either $g_0(L)=g^*(L_i)$ or $m_0(L)=m^*(L)$.
\end{ass}

\begin{ass}\label{C_ML}{Consistency of nuisance parameter estimators}\\
We have that $\norm{\hat{g}^c_k-g^*}_{P,2}=o_P(1)$, $\norm{\hat{m}^c_k-m^*}_{P,2}=o_P(1)$, $\norm{\hat{\alpha}_k\hat{G}_k-\alpha^*G^*}_{P,2}=o_P(1)$, and $\norm{\hat{\beta}_k\hat{M}_k-\beta^*M^*}_{P,2}=o_P(1)$
where $\hat{\alpha}_k\hat{G}_k$ and $\hat{\beta}_k\hat{M}_k$ are bounded above with probability approaching one.
\end{ass}

\begin{ass}\label{RC_ML}{Product rate conditions for nonparametric estimators}\\
$\sqrt{n}_k\norm{g^*-\hat{g}^c_k}_{P,2}\norm{m^*-\hat{m}^c_k}_{P,2}=o_P(1)$.\\
$\sqrt{n}_k\norm{\hat{\beta}_k\hat{M}_k}_{P,2} \norm{g^*-\hat{g}^c_k}_{P,2}=o_P(1)$,
$\sqrt{n}_k\norm{\hat{\alpha} \hat{G}_k}_{P,2} \norm{m^*-\hat{m}^c_k}_{P,2}=o_P(1)$,\\ $\sqrt{n}_k\norm{\hat{\alpha}\hat{G}_k}_{P,2}\norm{\hat{\beta}_k\hat{M}_k}_{P,2}=o_P(1)$.\\
If $g_0(L)\neq g^*(L)$ and $m_0(L)=m^*(L)$, we additionally require\\
$\sqrt{n}_k\norm{\bar{M}-M^*}_{P,2} \norm{m^*-\hat{m}^c_k}_{P,2}=o_P(1)$,$\sqrt{n}_k\norm{\bar{M}-M^*}_{P,2} \norm{\hat{\beta}_k\hat{M}_k}_{P,2}=o_P(1)$,
\\
$\sqrt{n}_k\norm{M^*-\hat{M}}_{P,2} \norm{m^*-\hat{m}^c_k}_{P,2}=o_P(1)$,
$\sqrt{n}_k\norm{M^*-\hat{M}}_{P,2} \norm{\hat{\beta}_k\hat{M}_k}_{P,2}=o_P(1)$.\\
If $m_0(L)\neq m^*(L)$ and $g_0(L)=g^*(L)$, we additionally require\\
$\sqrt{n}_k\norm{\bar{G}-G^*}_{P,2} \norm{g^*-\hat{g}^c_k}_{P,2}=o_P(1)$, $\sqrt{n}_k\norm{\bar{G}-G^*}_{P,2} \norm{\hat{\alpha}\hat{G}_k}_{P,2}=o_P(1)$,\\ $\sqrt{n}_k\norm{G^*-\hat{G}}_{P,2} \norm{g^*-\hat{g}^c_k}_{P,2}=o_P(1)$, $\sqrt{n}_k\norm{G^*-\hat{G}}_{P,2} \norm{\hat{\alpha}\hat{G}_k}_{P,2}=o_P(1)$.
\end{ass}

\begin{ass}\label{EP}{Empirical process conditions}
\begin{enumerate}[label = (\alph*)]
\item For a random subset $I$ of $\{1,...,n_k\}$ of size $n_k=n/k$, we have that the nuisance parameter estimator $\hat{\eta}^c_k$ belongs to a realisation set $\mathcal{N}$ with probability approaching one. The set $\mathcal{N}$ contains the limit $\eta^*$ of $\hat{\eta}^c_k$ and is constrained by the conditions below. 
\item $\hat{\alpha}$, $\hat{\beta}$, $\hat{G}(L)$ and $\hat{M}(L)$ are contained in (respective) uniformly bounded function classes $\mathcal{F}_\alpha$, $\mathcal{F}_\beta$, $\mathcal{F}_G$ and $\mathcal{F}_M$ with probability approaching 1. The sets $\mathcal{F}_\alpha$, $\mathcal{F}_\beta$, $\mathcal{F}_G$ and $\mathcal{F}_M$ include $\alpha^*$, $\beta^*$, $G^*(L)$ and $M^*(L)$; moreover, the space $\mathcal{T}=\mathcal{F}_\alpha \times \mathcal{F}_\beta \times \mathcal{F}_G \times \mathcal{F}_M$ is equipped with the product $L_2(P)$ semi-metric $d_2(\tau,\tilde{\tau})=\norm{\alpha-\tilde{\alpha}}_{P,2}+\norm{\beta-\tilde{\beta}}_{P,2}+\norm{G^*-\tilde{G}}_{P,2}+\norm{M^*-\tilde{M}}_{P,2}$.
\item For each $\eta\in\mathcal{N}$ and $j=1,2,3$, the function classes 
\begin{align*}
\mathcal{F}_j&=\left\{\psi_j(\cdot;\theta_0,\eta,\tau):\tau \in\mathcal{T}\right\}
\end{align*}
 are measurable with envelopes $F_j$, where $F_j\geq \sup_{f\in \mathcal{F}_j} |f|$  with $\norm{\mathcal{F}_j}_{P,q}< \infty$ for some $q\geq 2$ and $\max_{i\leq n}F(W_i)\leq \infty$.  Further, there exists a positive number $\xi\geq e$ and $\nu\geq 1$ such that the covering numbers satisfy the condition
\[\sup_Q \log N\left(\epsilon \norm{F_j}_{Q,2},\mathcal{F}_j,L_2(Q)\right)\leq \nu\log\left(\frac{\xi}{\epsilon}\right)\]
for all $0<\epsilon\leq 1$.
\item For $j=1,2,3$, we have that 
\[\sup_{\eta \in \mathcal{N},d_2(\tau,\tau^*)<\delta_{n_k}} \norm{\psi_j(\theta_0,\eta,\tau)-\psi_j(\theta_0,\eta^*,\tau^*)}_{P,2}=O_P(r^{(j)}_{n_k}) \]
where $\delta_{n_k}\to 0$, and $r^{(j)}_{n_k}$ is a rate that satisfies the restrictions $r^{(j)}_{n_k}\sqrt{\log\left(1/r^{(j)}_{n_k}\right)}=o(1)$ and $n_k^{-1/2}\log\left(1/r^{(j)}_{n_k}\right)=o(1)$. Furthermore, 
\[\sup_{\eta \in \mathcal{N}} \norm{\psi_j(\theta_0,\eta,\hat{\tau}_k)-\psi_j(\theta_0,\eta,\tau^*)}_{\mathbb{P}_{n,k},2}=o_P(1) \]

\end{enumerate}
\end{ass}

Assumption \ref{C_ML} requires that all estimators converge to a limit (which may not be the truth), and Assumption \ref{RC_ML} requires that certain combinations of the estimators of $g^*(L)$, $m^*(L)$, $G^*(L)$, $M^*(L)$, $\alpha^*$ and  $\beta^*$ converge sufficiently fast to a limit. We use the $L_2$ distance to define a measure of consistency here, since this is what naturally arises when bounding terms in the expansion using the Cauchy-Schwarz inequality. Regarding the conditions on $\norm{\bar{M}-M^*}_{P,2}=$ and $\norm{\bar{G}-G^*}_{P,2}$, we note that these loosely require a regression function conditional on a generated regressor to converge to the function conditional on the limit of the regressor. 

We place assumptions on $\hat{\alpha}_k\hat{G}_k$ and $\hat{\beta}_k\hat{M}_k$ converging to zero, rather than on the $\hat{\alpha}_k$, $\hat{\beta}_k$, $\hat{G}_k$ and $\hat{M}_k$ individually converging. This is in light of the aforementioned issue that the denominators of 
$\hat{\alpha}_k$ and $\hat{\beta}_k$ may themselves converge to zero when nuisance parameters are consistently estimated, whilst it suffices that the products taken with $\hat{G}_k$ and $\hat{M}_k$ converge. This issue also arises in previous work on doubly robust inference \citep{van2014targeted,benkeser2017doubly}; a formal analysis of the impact on  moderate-sample performance is an important topic for future work. In the Appendix, we show that when $m^*(L)\neq m_0(L)$, the convergence rate of $\hat{\alpha}$ will typically depend on $\norm{G^*-\hat{G}}_{P,2}$ and $\norm{g^*-\hat{g}}_{P,2}$, and its rate will typically be dictated by whichever converges slower. Assumptions on the convergence of $\bar{M}(L)$ to $M^*(L)$ (and $\bar{G}(L)$ to $G^*(L)$) have been made in previous work \citep{benkeser2017doubly}, and are generally plausible if $\hat{m}(L)$ and $\hat{g}(L)$ converge sufficiently fast to their limits. 

Assumption \ref{EP} places restrictions on the complexity of $\tau^*$, but does not restrict the complexity of $\hat{\eta}(I^c_k)$ or $\eta^*$. Essentially, we require $\tau^*$ to belong to a Vapnik–Chervonenkis (VC) class (such that it has a covering number bounded by a polynomial in $1/\epsilon$); a class that is of VC type is also Donsker \citep{van1996weak}. Conditional on $I^c_k$, these entropy conditions are met for $\alpha$ and $\beta$ (and their estimators) since these are scalar parameters in parametric models and thus the corresponding function classes typically have uniformly bounded entropy. Also, $M^*(L)$ and $G^*(L)$ are estimated non-parametrically, but they correspond to univariate nonparametric regressions (where the entropy can again be bounded). Simpler proofs using empirical process results could be constructed if one were to assume that the envelopes for the considered function classes go to zero and that $J(1,\mathcal{F},L_2)<\infty$; however, in general the function class (and hence the uniform entropy integrals) will depend on $n$ via $\hat{\eta}(I^c_k)$, so we take a different approach below, looking at specific properties of VC classes. In order to apply the local maximal inequality of \citet{chernozhukov2014gaussian}, we place a restriction on covering numbers, but we believe that a similar result could be developed via a restriction on the bracketing number and a bound on the bracketing entropy integral \citep{van1996weak}. See for example Theorem 3 of \citet{kennedy2019nonparametric}. This would apply to smooth functions and Sobolev classes; for example, the bracketing entropy integral converges so long as the H\"older exponent is greater than half of the covariate dimension. We note that Assumption \ref{EP} could be avoided using a triple-splitting procedure, where $\eta^*$ is estimated in one split, $\tau^*$ is estimated on the second and inference on $\theta_0$ is performed in the third.

Our key theorem is as follows:
\begin{theorem}\label{key_theo}
Under Assumptions \ref{con_ML}-\ref{EP},
\begin{align*}
\sqrt{n_k}\mathbb{P}_{n,k} \psi_1(\theta_0,\hat{\eta}^c_k,\hat{\tau}_k)&=\sqrt{n_k}\mathbb{P}_{n,k} \psi_1(\theta_0,\eta^*,\tau^*)\\&\quad-I\{m^*(L)=m_0(L)\}\sqrt{n_k}\mathbb{P}_{n,k}\psi_2(\theta_0,\eta^*,\tau^*)\\&\quad-I\{g^*(L)=g_0(L)\}\sqrt{n_k}\mathbb{P}_{n,k}\psi_3(\theta_0,\eta^*,\tau^*)\\&\quad+o_P(1)
\end{align*}
\end{theorem}
Let us define 
\[T_{n,k}=\frac{\sqrt{n}_k \mathbb{P}_{n,k}\psi^*(\theta_0,\hat{\eta}^c_k,\hat{\tau}_k)}{\left[\mathbb{P}_{n,k}\psi^*(\theta_0,\hat{\eta}^c_k,\hat{\tau}_k)^2-\{\mathbb{P}_{n,k}\psi^*(\theta_0,\hat{\eta}^c_k,\hat{\tau}_k)\} ^2\right]^{1/2}}\]
Then Theorem \ref{key_theo} can be extended to show asymptotic type I error control. 
\begin{corollary}\label{key_cor}
Suppose that Assumptions \ref{con_ML}-\ref{EP} hold, and furthermore that $0<P\{\psi_1(\theta_0,\eta^*,\tau^*)-\psi_2(\theta_0,\eta^*,\tau^*)\}^2<\infty$ and $0<P\{\psi_1(\theta_0,\eta^*,\tau^*)-\psi_3(\theta_0,\eta^*,\tau^*)\}^2<\infty$then 
\[\sup_{t\in \mathbb{R}}|Pr(T_{n,k}\geq t)-\Phi(t)|\to 0\]
\end{corollary}
This result shows that our proposed test continues to be valid so long as at least two nuisances are consistently estimated; we need either
\begin{enumerate}[label=(\roman*)]
\item $\norm{g_0-\hat{g}^c_k}_{P,2}\norm{m_0-\hat{m}^c_k}_{P,2}=o_{P}(n^{-1/2})$ or
\item $\norm{g_0-\hat{g}^c_k}_{P,2}\norm{G^*-\hat{G}_k}_{P,2}=o_{P}(n^{-1/2})$ or
\item $\norm{m_0-\hat{m}^c_k}_{P,2}\norm{M^*-\hat{M}_k}_{P,2}=o_{P}(n^{-1/2})$.
\end{enumerate}
When either $g_0(L)$ or $m_0(L)$ cannot be consistently estimated, we continue to obtain root-$n$ inference by transferring our assumptions to a different nuisance parameter (either $G^*(L)$ or $M^*(L)$). As stated before, these nuisances should be less difficult to estimate because they are univariate regressions. Nevertheless, since all nuisances are dependent on machine learning methods, we will see in Section \ref{kernel_sec} that convergence of both $\hat{G}(L)$ and $\hat{M}(L)$ can depend in a complex way on the convergence rates of $\hat{g}(L)$ and $\hat{m}(L)$.

The above result can be used to justify type I error control of our test. Consider a series of tests $(T_n)^\infty_{n=1}$; let $\mathcal{P}$ denote a set of laws satisfying the null hypothesis of interest and also certain regularity conditions (including smoothness or sparsity conditions). We focus predominantly on \textit{pointwise asymptotic error control}:
\[\sup_{P\in \mathcal{P}} \limsup_{n\to \infty} Pr_{P}(T_n=1)\leq \alpha\]
for a level $\alpha \in (0,1)$. This is in contrast with \textit{uniform asymptotic error control}:
\[\limsup_{n\to \infty} \sup_{P\in \mathcal{P}}  Pr_{P}(T_n=1)\leq \alpha\]
which generally gives much stronger guarantees; see e.g. \citet{lehmann20226testing} for further details. We note that in the case that both nuisances are consistently estimated, then uniform guarantees should be feasible given a slight adjustment of the results for testing in \citet{shah2020hardness} and for estimation/interval-construction in \citet{chernozhukov2018double}. Given that the focus of this paper is on doubly robust inference, we have chosen not to pursue this here. However, in light of the results on regularity and superefficiency in the following section, we would conjecture that pointwise results cannot be made uniform when only one of the two nuisances is consistently estimated (unless parametric working models are used).

Under Assumptions \ref{con_ML}-\ref{EP}, the proposed (unnormalized) doubly robust score statistic $U_n(\theta_0) = K^{-1} \sum_{k=1}^{K}\tilde{U}_{n_k}(I_k,I_k^c)$ is asymptotically linear with influence function\\ $\psi^*(W;\theta_0,\eta^*,\tau^*) = \psi_1(W;\theta_0,\eta^*,\tau^*) - I\{m^*(L)=m_0(L)\}\psi_2(W;\theta_0,\eta^*,\tau^*) - I\{g^*(L)=g_0(L)\}\psi_3(W;\theta_0,\eta^*,\tau^*).$ For the remainder of this section, we will sometimes omit the dependence of $\psi^*(W;\theta_0,\eta^*,\tau^*)$ on $\eta^*,\tau^*$ to simplify notation.
The following assumptions allow us to conclude the stronger statement that $U_n(\theta)$ is asymptotically linear as a function of $\theta$; let $\Theta$ denote the parameter space for $\theta$, which is assumed to be bounded. In the following assumptions, we now index nuisance parameters by $\theta$ to make explicit which ones are dependent on the target parameter. 

\begin{ass}{Uniform convergence of the scores}\label{uniEP}
    \item $\sup_{\theta \in \Theta} P \{ \hat{\psi}^*(\theta)-\psi^*(\theta) \}^2 = o_P(1).$
\end{ass}
\begin{ass}{Uniform convergence of the remainder terms}\label{uni_ML}
    \begin{enumerate}[label=(\roman*)]
        \item $\sup_{\theta \in \Theta} \norm{g_0-\hat{g}}_{P,2}\norm{m_0(\theta)-\hat{m}^c_k(\theta)}_{P,2}=o_{P}(n_k^{-1/2})$ or
        \item $ \sup_{\theta\in \Theta} \norm{g_0-\hat{g}^c_k}_{P,2}\norm{G^*(\theta)-\hat{G}_k(\theta)}_{P,2}=o_{P}(n_k^{-1/2})$ or
        \item $\sup_{\theta\in \Theta} \norm{m_0(\theta)-\hat{m}^c_k(\theta)}_{P,2}\norm{M^*(\theta)-\hat{M}_k(\theta)}_{P,2}=o_{P}(n_k^{-1/2})$
\end{enumerate}
\end{ass}
	
We now consider the proposed estimator $\hat{\theta}_n$ such that $U_n(\hat{\theta}_n) = 0$. Under the additional conditions \ref{uniEP} and \ref{uni_ML}, we have that $n^{1/2} \{U_n(\theta)-P\psi^*(\theta)\}$ converges weakly to a Gaussian process with mean zero and covariance function $\rho(\theta_1,\theta_2)=P\{\psi^*(\theta_1)\psi^*(\theta_2)\}$. By the delta method, $\hat{\theta}_n$ is asymptotically linear with influence function $-\{\frac{\partial}{\partial\theta_0} P \psi^*(\theta_0)\}^{-1} \psi^*(W;\theta_0)$ \citep{van1996weak}.

\subsection{Regularity and superefficiency}
In this section, we show that the doubly robust sampling distribution conferred by Theorem \ref{key_theo} comes at the cost of regularity when either $\hat{g}(L)$ or $\hat{m}(L)$ is not consistent for the true nuisance function. To study regularity of the proposal, we characterize the behavior of the estimator obtained by inverting the score test under a sequence of local alternatives. 

To simplify the exposition below, we consider the case where the outcome and exposure are known to follow a multivariate Gaussian distribution conditionally given the covariates $L$. The distribution of the covariates remains unconstrained (nonparametric). Hence, the joint density of $W=(Y,A,L)$ is
	$$p_{\theta,\eta}(W)=\phi\{Y-A\,\theta-m(L)\}\,\phi\{A-g(L)\}\,p_L(L)$$
where $\phi(z) = (2\pi)^{-1/2}\exp(-z^2/2)$ and $p_L$ is the unspecified joint density of the covariate vector $L$.
	
Let $\{\theta_t,\eta_t(L)\} = \{\theta_0 + t s_\theta, g_0(L) + t s_g(L), m_0(L) + t s_m(L) \}$ denote a parametric submodel in the direction $s=\{s_\theta,s_m(L),s_g(L)\}$ and passing through the population parameter value $(\theta_0,\eta_0)$ at $t=0$. The scores of $p_{\theta_t,\eta_t}$ (with respect to $t$) at $t=0$ are given by the linear operator $s \mapsto \mathbb{B} s$ defined by
$$ \mathbb{B} s(W) = \{Y-A\theta_0-m_0(L)\} \{As_\theta + s_m(L)\} + \{A -g_0(L)\} s_g(L) . $$
A more general analysis for density functions other than $\phi$ will exhibit a similar form.
	
We now replace $\{\theta_t,\eta_t(L)\}$ by the local parameter sequence $\{\theta_{n^{-1/2}}, \eta_{n^{-1/2}}(L)\} = \{\theta_0 + n^{-1/2} s_\theta, g_0(L) + n^{-1/2} s_g(L), m_0(L) + n^{-1/2} s_m(L) \}$ and let $P_n$ denote the corresponding local probability measures. By arguments identical to those in \cite{van2000asymptotic} and \cite{banerjee2005likelihood}, the log likelihood ratios $\Lambda_n = \log P^n - \log P^n_n$ are locally asymptotically normal with
$$ \Lambda_n = n^{1/2} \mathbb{P}_n \mathbb{B}s - \frac{1}{2} P(\mathbb{B}s)^2 + o_{P}(1) \overset{P}{\rightsquigarrow} N\left(- \frac{1}{2} P(\mathbb{B}s)^2, P(\mathbb{B}s)^2 \right) $$
where $\overset{P}{\rightsquigarrow}$ denotes weak convergence with respect to $P$ and the $o_{P}(1)$ term is also $o_{P_n}(1)$. See chapter 7 of \cite{van2000asymptotic} for a detailed account of local asymptotic normality.

We now consider regularity of the proposed estimator. Recall that an estimator $T_n$ of $\theta_0$ is (locally) regular at $P$ if the limiting distribution of $n^{1/2}(T_n - \theta_{n^{-1/2}})$ with respect to sampling under $P_n$ does not depend on $s$. Standard definitions of asymptotic efficiency restrict attention to regular estimators in order to rule out so-called superefficient estimators \citep{van2000asymptotic}.

The next result characterizes the regularity of $\hat{\theta}_n$ through its sampling distribution under sequences of local parameters.
	
\begin{theorem}\label{reg_theo}
	Suppose that the likelihood of an observation is
	$$ p_{\theta,\eta}(W) = \phi\{Y-A\,\theta-m(L)\}\,\phi\{A-g(L)\}\,p_L(L) $$
	where $\phi(z) = (2\pi)^{-1/2}\exp(-z^2/2)$ and $p_L$ is the unspecified joint density of the covariate vector $L$. If Assumptions \ref{con_ML}-\ref{uni_ML} hold, then
	$$ n^{1/2}(\hat{\theta}_n-\theta_n) \overset{P_n}{\rightsquigarrow} N(\mu_s, \tau^2_*) $$
	where $\{\theta_{n^{-1/2}}, \eta_{n^{-1/2}}(L)\} = \{\theta_0 + n^{-1/2} s_\theta, g_0(L) + n^{-1/2} s_g(L), m_0(L) + n^{-1/2} s_m(L) \}$ for $s=\{s_\theta,s_m(L),s_g(L)\}$. The asymptotic variance is
	$$ \tau^2_* = P \left[ \left\{\frac{\partial}{\partial\theta_0} P \psi^*(\theta_0)\right\}^{-1} \psi^*(\theta_0) \right]^2. $$
	With some abuse of notation, the asymptotic mean $\mu_s$ is
	$$ \mu_s = \begin{cases}
		0 & \text{if } m^*(L)=m_0(L), g^*(L)=g_0(L) \\
		\frac{P\{(A-g^*-M^*)s_m\}}{P\{(A-g^*-M^*)A\}}
		 & \text{if } m^*(L)=m_0(L), g^*(L) \neq g_0(L) \\
		 \frac{P\{(Y-A\theta_0-m^*-G^*)s_g\}}{P\{(A-g_0)A\}} & \text{if } m^*(L) \neq m_0(L), g^*(L)=g_0(L)\,.
	\end{cases} $$
\end{theorem}
	
This theorem follows by an application of Le Cam's third lemma \citep{van2000asymptotic}. A consequence of this result is that $\hat{\theta}_n$ is a regular asymptotically linear estimator if both nuisance regressions are consistently estimated. However, if $g^*(L)\neq g_0(L)$, then $\hat{\theta}_n$ remains asymptotically linear but is only locally regular with respect to paths $s=\{s_\theta,0,s_g(L)\}$ that do not vary $m_0(L)$. Similarly, when $m^*(L)\neq m_0(L)$, the estimator $\hat{\theta}_n$ is locally regular for paths $s=\{s_\theta,s_m(L),0\}$ that do not vary $g_0(L)$. The asymptotic results in the previous sub-section may thus give a poor approximation under certain data-generating mechanisms, and erratic behaviour is expected to be more severe as the degree of misspecification is larger. Note that other examples of doubly robust asymptotically linear estimators such as \cite{benkeser2017doubly} focus on parameters in nonparametric models where only one influence function is possible for regular asymptotically linear estimators (see the following subsection). Hence, the result here is non-trivial for the semiparametric model under consideration. 

Recently, \citet{benkeser2020superefficient} have proposed superefficient estimators of average treatment effect parameters by replacing the propensity score with an alternative regression. Proposals such as these forsake regularity of their estimators by their construction in order to achieve lower overall mean squared error. Relating this to the above theorem, note first that the semiparametric efficiency bound under the partially linear model with homoscedastic errors is proportional to $1/\mathbb{E}\{var(A|L)\}$. If one considers a more general class of non-regular (and non-asymptotically linear) estimators of $\theta_0$ then it is possible to improve this in parts of the parameter space. When $g^*(L)\neq g_0(L)$, the influence function for the proposed estimator is scaled by $1/\mathbb{E}[\{(A-g^*(L)-M^*(L)\}A]$;  then $g^*(L)$ can be chosen to deliver an estimator with asymptotic variance smaller than the semiparametric efficiency bound but greater than $1/var(A)$ e.g. by letting it not depend on $L$. However, in light of Theorem \ref{reg_theo}, we advocate for the more cautious approach of attempting to model both $m_0(L)$ and $g_0(L)$ correctly and viewing the results of Theorem \ref{key_theo} as offering some additional protection against possible misspecification.

\subsection{Doubly robust inference for the variance-weighted average treatment effect}

Throughout this article, we have assumed that the semiparametric model of interest is correctly specified e.g. that (\ref{plm}) holds. When this is not the case, then so long as $g_0(L)$ is consistently estimated, then our proposed estimator of the parameter in the partially linear model converges to 
\begin{align}\label{np_estimand}
\theta^*_0= \frac{\mathbb{E}[\{A-g_0(L)\}Y]}{\mathbb{E}[\{A-g_0(L)\}A]}
\end{align}
as discussed e.g. in \citet{robins2008higher}. When $A$ is binary, and $L$ is sufficient to account for confounding, then the above equals the so called \textit{overlap-weighted treatment effect} \citep{crump2006moving}. 
However, under a nonparametric model for the observed data distribution, there is only one regular and asymptotically linear estimator of (\ref{np_estimand}), which has the influence function:
\begin{align}\label{if_np}
\frac{\{A-g_0(L)\}[Y-\theta^*_0\{A-g_0(L)\}-\mathbb{E}(Y|L)]}{\mathbb{E}[\{A-g_0(L)\}^2]}.
\end{align}
Indeed, one can show that the influence function of the standard doubly robust estimator $\hat{\theta}$ in (\ref{dr_est}) is equal to (\ref{if_np}) plus a term
\[\frac{\{A-g_0(L)\}\{\mathbb{E}(Y-\theta^*_0A|L)-m^*(L)\}}{\mathbb{E}[\{A-g_0(L)\}^2]},
\]
with a drift term now equal to $P[\{g_0(L)-\hat{g}(L)\}\{\mathbb{E}(Y-\theta^*_0A|L)-\hat{m}(L)\}]$. Then since $\hat{m}(L)$ does not converge to $\mathbb{E}(Y-\theta^*_0A|L)$, the latter will usually fail to be $o_P(n^{-1/2})$ when $g_0(L)$ is estimated data-adaptively \citep{whitney2019}.
The influence function for our proposed estimator is equal to (\ref{if_np}) plus a term
\[\frac{\{A-g_0(L)\}[\mathbb{E}(Y-m^*(L)|L)-\mathbb{E}\{Y-m^*(L)|g_0(L)\}]}{\mathbb{E}[\{A-g_0(L)\}^2]},
\]
with a drift term equal to
\begin{align*}
&P[(g_0-\hat{g})(G^*-\hat{G})\}+P\{(g_0-\hat{g})(m^*-\hat{m})\}
\end{align*}
which is $o_P(n^{-1/2})$ under the conditions of Theorem \ref{key_theo} above. Therefore unlike the standard doubly robust estimator, our proposal yields an asymptotically linear (although non-regular) estimator of $\theta^*_0$ when restriction (\ref{plm}) fails.

\subsection{Results for estimators of $M(L)$ and $G(L)$}\label{kernel_sec}

In this section, we explore the properties of the Nadaraya-Watson estimators of the additional nuisance parameters and obtain novel results on their rates of convergence; given that similar results are available for both $\hat{M}(L)$ and $\hat{G}(L)$, in what follows we will focus on the former. The following theorem characterises the convergence rate when the estimators of $g_0(L)$ and $m_0(L)$ are taken from an auxiliary sample. \citet{mammen2012nonparametric} and \citet{delaigle2016nonparametric} give comparable results in settings where either the regressors and/or the outcome in a kernel regression problem depend on estimated nuisance parameters (see also \citet{kennedy2017nonparametric}). In these papers, results are given for specific nonparametric estimators of the nuisances. Our results allow for arbitrary estimators or $g_0(L)$ and $m_0(L)$ by incorporating sample splitting.
\begin{theorem}\label{kern_theo}
For a random subset $I$ of $\{1,...,n_k\}$ of size $n_k=n/k$, suppose we have that the nuisance parameter estimator $\hat{\eta}(I^c_k)$ obtained on $I^c_k$ belongs to the realisation set $\mathcal{N}$ with probability approaching one. The set $\mathcal{N}$ contains $\eta^*$ and satisfies Assumption \ref{EP} required for Theorem \ref{key_theo}. For an $x\in \mathcal{X}$, where $\mathcal{X}$ denotes the support of $m^*(L)$, suppose that 
\begin{enumerate}[label=(\alph*)]
\item For a chosen bandwidth $h$, $h\to0$, $nh^3\to\infty$ and $n\to\infty$.\label{C_K1}
\item $K$ is $\vartheta$th-order kernel where, for some $\nu<\infty$, $|K(s)-K(s')|\leq \nu|s-s'| $ for all $s, s' \in \mathbb{R}^1$. \label{C_K2}
\item The support $\mathcal{X}$ is compact. Also, $f_{m}(x)=\partial P(m^* \leq x)/\partial x$ is continuous in $x$ and bounded away from zero.\label{C_K3}
\item $M^*(x)$ is $\vartheta$-times continuously differentiable. The conditional density of $A-g^*(L)$ given $m^*(L)=x$ is continuous w.r.t. $x$.\label{C_CD}
\item The kernel $K$ belongs to a class $\mathcal{F}(K)=\{K(\frac{x-\cdot}{h}):x\in \mathcal{X},h>0\}$ which is of VC-type with $\xi\geq e$, $\nu\geq 1$ and uniformly bounded envelope $F(K)<\infty$. \label{C_ent}
\item The mean-square misspecification error satisfies $\norm{g_0-g^*}_{P,2}=O(1)$. \label{ms_b}
\item For each $j=1,2$ and each $k$, 
\[\sup _{x\in \mathcal{X},\eta \in \mathcal{N}}\norm{\varphi_j(x,h,\eta)-\varphi_j(x,h,\eta^*)}_{P,2}=O_P(\sqrt{h}t^{(j)}_{n_k})\] where $t^{(j)}_{n_k}$ is a rate that satisfies $\sqrt{h}t^{(j)}_{n_k} \sqrt{\log \left(\frac{1}{\sqrt{h}t^{(2)}_{n_k}}\right)}\to 0$ and $\frac{1}{\sqrt{n_k}} \log \left(\frac{1}{\sqrt{h}t^{(2)}_{n_k}}\right)\to 0$.  \label{C_con}

\end{enumerate}
Then 
\begin{align}\label{ktres1}
|\hat{M}(x)-M^*(x)|=O_P\left(h^\vartheta +\zeta_g+h^{-1}\zeta_m\right)
\end{align}
and 
\begin{align}\label{ktres2}
\mathbb{E}[\{\hat{M}(x)-M^*(x)\}^2]=O\left(h^{2\vartheta}+\zeta^2_g+h^{-2}\zeta^2_m\right),
\end{align}
where 
\[\norm{g^*-\hat{g}^c_k}_{P,2}=O_P(\zeta_g)\textrm{ and }\norm{m^*-\hat{m}^c_k}_{P,2}=O_P(\zeta_m).\]
\end{theorem}

Here we briefly discuss the assumptions. Conditions \ref{C_K1}-\ref{C_CD} are standard in the kernel regression literature. We require conditions  \ref{C_ent} and \ref{C_con} in order to apply results on empirical processes. The former concerns the complexity of the terms which comprise the kernel regression estimator (and their estimators). Importantly, the entropy of the relevant function classes will be controlled for fixed $\hat{\eta}(I^c_k)$ and $\eta$, such that we do not need to restrict the entropy of $\hat{g}(L)$ and $\hat{m}(L)$ here. Condition \ref{C_ent} is satisfied for many common choices of kernel, and means that $\mathcal{F}(K)$ is of VC-type and therefore that the uniform entropy condition \ref{C_ent} is satisfied \citep{gine2002rates}. 

We now unpack the implications of Theorem \ref{kern_theo}. Firstly, the behaviour of $\hat{M}(L)$ in general depends on both $\hat{m}(L)$ and $\hat{g}(L)$, such that even though $\hat{g}(L)$ does not have to converge to the truth, we still rely on it to converge relatively quickly to some limiting value. This situation is preferable to one where $\hat{g}(L)$ converges to the truth but the rate is so slow that the usual Cauchy-Schwarz bound fails to be $o_P(n^{-1/2})$. As a consequence, if either $g_0(L)$ or $m_0(L)$ is especially difficult to estimate, these results suggest that one might prefer an inconsistent estimator that converges quickly to some $g^*(L)$ rather than one that converges to the truth, at least from the perspective of high quality estimation of $M^*(L)$ or $G^*(L)$. Nevertheless, the pointwise nature of the asymptotic results for doubly robust inference suggest that for certain data generating processes, fast convergence of $\hat{M}(L)$ or $\hat{G}(L)$ will not necessarily translate into good size/coverage of the resulting tests or confidence intervals. This is borne out in the simulations (see Experiment 3) in the following sections, where we see that certain types of misspecification can result in worse performance than letting $\hat{m}(L)$ and $\hat{g}(L)$ converge slowly to the truth. 

Secondly, there are immediate implications for how to combine cross-fitting with doubly robust inference. In order to avoid placing Donsker conditions on $\hat{g}(L)$ and $\hat{m}(L)$ and their limits, we should obtain them in a separate sample to the one used to obtain $\hat{M}(L)$ and $\hat{G}(L)$. However, in the proof of Theorem \ref{key_theo} where sample splitting is required, empirical process conditions are placed on the latter estimators, given that they are univariate nonparametric estimators and that $\hat{g}(L;I^c_k)$ and $\hat{m}(L;I^c_k)$ are considered fixed (conditional on the auxiliary sample). 

Finally, the optimal bandwidth for $\hat{M}(L)$ will now depend on the rate of convergence of $\hat{m}(L)$. Note that 
the usual variance term of order $1/\sqrt{n_kh}$ that appears in the expansion of a kernel estimator is $o(h^{-1}\zeta_m)$ in expression \eqref{ktres1}. To see this, suppose that $h=n^{-\lambda}$ and $\zeta_m=n^{-\upsilon}$ for some $0<\lambda<1$ and $0<\upsilon\leq1/2$ respectively. Then, $1/\sqrt{n_kh}=o(h^{-1}\zeta_m)$ if $2\upsilon-1<\lambda$ which holds for all $\lambda$ and $\upsilon$. It furthermore follows that the optimal choice of bandwidth for balancing squared bias and variance is $h\sim \zeta_m^{1/(2\vartheta-1)}$. Hence, letting $h$ tend more slowly to zero is advantageous when $\hat{m}(L)$ converges slowly.

As pointed out by a reviewer, it may be possible to improve the term $h^{-1}\zeta_m$ in the rate above, under strengthened conditions along the lines of Theorem 1 in \citet{mammen2012nonparametric}. Specifically, if one assumes that $E\{A|m^*(L)\}=g_0(L)$, then we would conjecture that the term could be improved to to $\zeta_m$. We would not typically want to enforce this condition, given that we have not even assumed that $g_0(L)$ is consistently estimated by $\hat{g}(L)$. It may be possible to weaken this to a smoothness condition on the difference $E\{A|m^*(L)\}-g_0(L)$ whilst maintaining the improved rate; see Assumption 4 in \citet{mammen2012nonparametric} for example. However, we leave this to further work, since we do not find such a condition intuitive in the doubly robust inference set-up.

\section{Simulation studies}

In order to judge how well the methods are expected to perform in practice, we conducted three simulation experiments. 

\subsection*{Experiments 1 \& 2}

We considered four score tests of the null hypothesis that $\theta=0$. The first, `PS' (propensity score), is based on the score statistic $n^{-1/2}\sum^n_{i=1}\{A_i-\hat{g}(L_i)\}Y_i$ after scaling by its empirical standard deviation. The second, `OR' (outcome regression), is based on the statistic $\sqrt{n}\mathbb{P}_n[A_i\{Y_i-\hat{m}(L_i)\}]$ (after scaling); neither of the first two tests possess the \textit{Neyman orthogonality} property discussed in \citet{chernozhukov2018double}, and hence would not be expected to posses their advertised size when data adaptive estimators of the nuisance parameters are used. We also considered a `GCM' test \citep{shah2020hardness}, based on the orthogonalised statistic $n^{-1/2}\sum^n_{i=1}\{A_i-\hat{g}(L_i)\}\{Y_i-\hat{m}(L_i)\}]$, in addition to our proposal `DR-GCM' given in Section \ref{proposal}. Although it is not generally required \citep{shah2020hardness}, our implementation of the `GCM' test uses cross-fitting, as this was also done for the proposal and we wanted to ensure a fair comparison of methods. One would expect the `GCM' test to posses its advertised size when both nuisance parameter estimators are consistent, but not otherwise, in contrast to our proposal. In order to construct the tests, all statistics given above were scaled by the empirical standard deviation of the relevant scores.

The first experiment was closely related to the setting in \citet{benkeser2017doubly}. The first covariate $L_1$ was generated from a $U(-2,2)$ distribution, whilst the second covariate $L_2$ and exposure were both binary with respective expectations 0.5 and $expit\{-L_1+2L_1L_2\}$. The outcome $Y$ was simulated from a $\mathcal{N}(-L_1+2L_1L_2,1)$ distribution. We wished to evaluate how the aforementioned tests would perform in 3 different settings: 1) when both nuisance parameter estimators $\hat{g}(L)$ and $\hat{m}(L)$ are consistent, 2) when only the estimated propensity score $\hat{g}(L)$ is consistent, and 3) when only the estimated outcome regression estimate is consistent. The parameter that was consistently estimated was obtained using the Super Learner \citep{van2007super}, whilst the inconsistently estimated parameter was obtained via $\ell_1-$penalised maximum likelihood with an omitted interaction term. The Super Learner library incorporated a sample average, a generalised linear model with an interaction term, a kernel $k$-nearest neighbours algorithm, random forests and several smoothing spline-based estimators with varying choices of penalty parameter. Nadaraya–Watson kernel regression estimators were used to estimate  $G^*(L)$ and $M^*(L)$. Tuning parameters were selected using cross-validation; we acknowledge that this may lack theoretical justification, particularly in light of the complex dependence of $G^*(L)$ and $M^*(L)$ on the other nuisance parameters. Implementation of optimal choices of tuning parameters with nonparametrically generated outcomes and covariates has not been studied, as far as we are aware; we note that  cross-validation was seen to perform well in \citet{benkeser2017doubly}.

\begin{figure}[h]
  \caption{First experiment (all nuisances estimated consistently).  \textcolor{black}{Black} dots represent `PS'; \textcolor{red}{red} squares represent `OR'; \textcolor{green}{green} triangles represent `DML'; \textcolor{blue}{blue} triangles with represent `DR-GCM'.}
  \centering
    \includegraphics[width=0.55\textwidth]{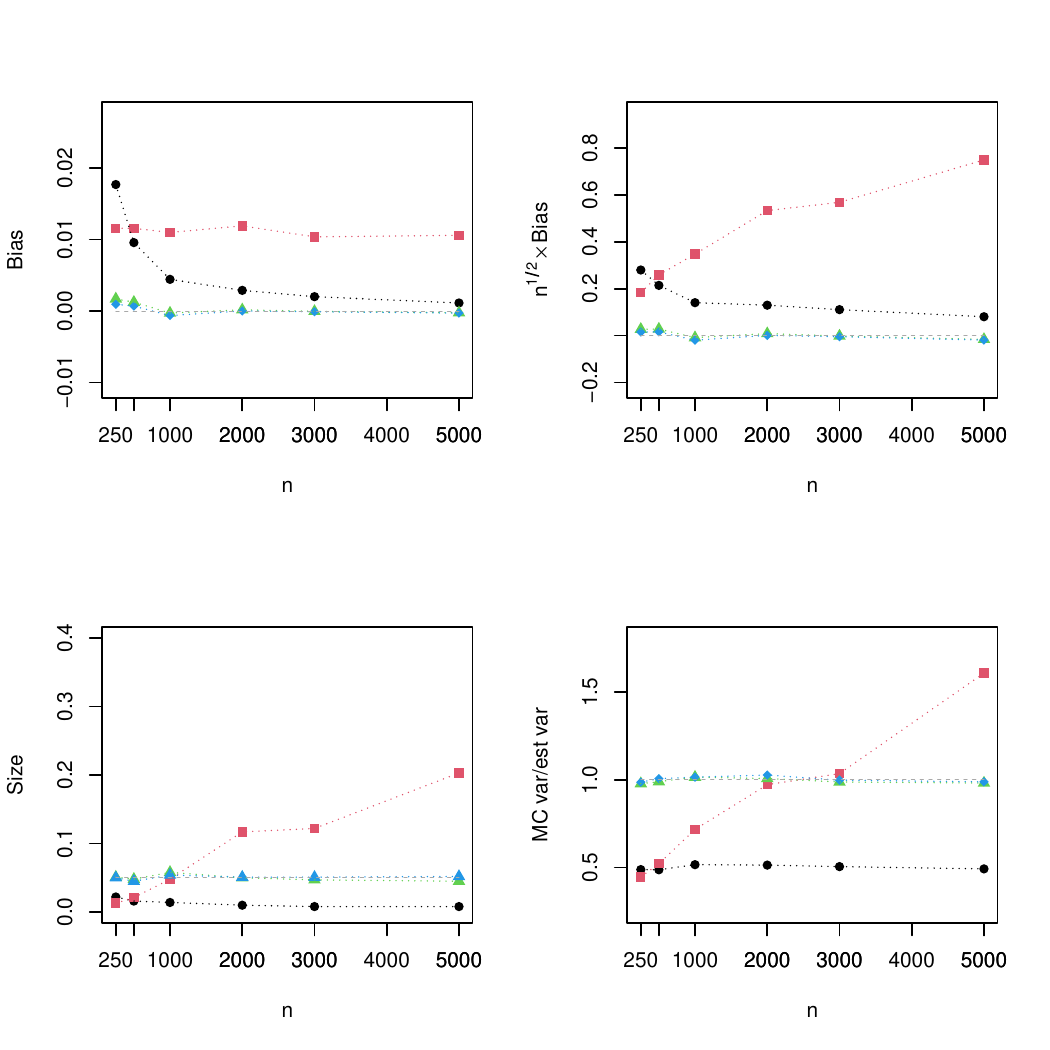}
\end{figure}

\begin{figure}
  \caption{First experiment (propensity score estimated correctly). \textcolor{black}{Black} dots represent `PS'; \textcolor{green}{green} triangles represent `GCM'; \textcolor{blue}{blue} triangles with represent `DR-GCM'.}
  \centering
    \includegraphics[width=0.55\textwidth]{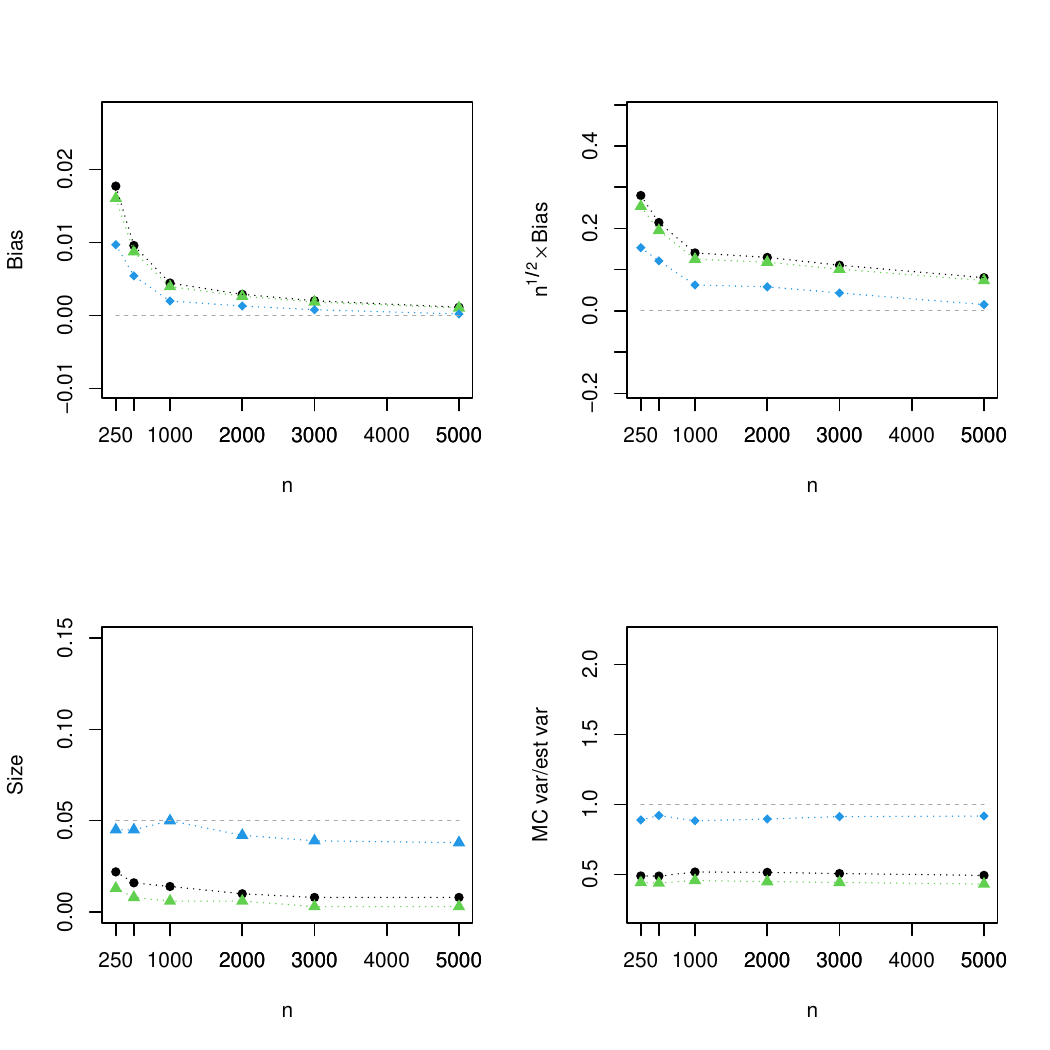}
\end{figure}

\begin{figure}
  \caption{First experiment (outcome model estimated correctly). \textcolor{red}{Red} squares represent `OR'; \textcolor{green}{green} triangles represent `GCM'; \textcolor{blue}{blue} triangles with represent `DR-GCM'.}
  \centering
    \includegraphics[width=0.55\textwidth]{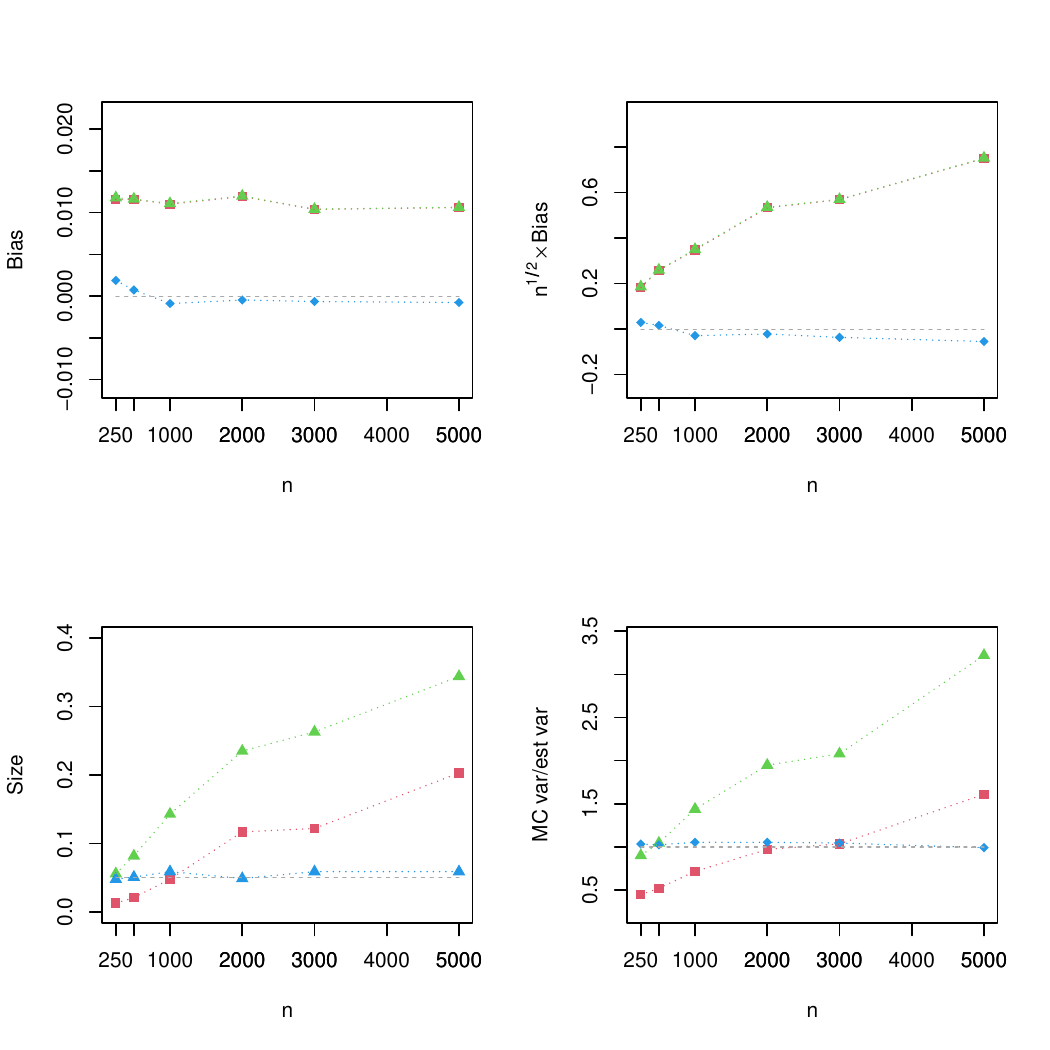}
\end{figure}

The second experiment considered a different type of misspecification. The covariates $L_1$ and $L_2$ were both generated from a uniform distribution over the interval $[0,1]$. The binary exposure had expectation $expit\{-1.5+3(1+e^{-20(L_1-0.5)})^{-1}+L_2\}$ and the outcome was generated from a $\mathcal{N}(3(1+e^{-20(L_1-0.5)})^{-1}+L_2,1)$ distribution. Inconsistently estimated nuisance parameters were again estimated via $\ell_1-$penalised maximum likelihood with only main effect terms. The Super Learner included the same candidate learners as in the previous experiment, and was again believed to be consistent; random forests have been seen to perform well in similar settings \citep{friedberg2020local,cui2019bias}, although the strong linear signal could be considered adversarial. All experiments were carried out at sample sizes of 250, 500, 1,000, 2,000, 3,000 and 5,000, and at each sample size repeated 1,000 times. Five-fold cross-fitting was used in the construction of each of the tests. 

The results of experiment 1 can be seen in Figures 1-3. Against sample size, we plot bias of the test statistic, $n^{1/2}\times$bias, the size of the test, and the Monte Carlo variance of the score statistic scaled by its average estimated variance (MC var/est var) based on the empirical variance of the score in each simulation. This ratio indicates whether the estimated variance used to scale the test is too high (a value less than one) or too low (a value greater than one).  The `OR' and `PS' tests failed to obtain their advertised rejection rate even when nuisance parameters were consistently estimated due to the slowly-converging bias inherited from these Super Learner-based estimators. The `GCM' test performed well when both nuisances were accurately estimated, but could display erratic behaviour otherwise. This was especially the case when only the outcome regression estimate was consistent, since the `GCM' test statistic appears to inherit the plug-in bias of $\hat{m}(L)$. This led to a rejection rate that increases with sample size, and was over 30\% at $n=5,000$. In contrast, the `DR-GCM' had close to its advertised size across all three settings. The results of the second experiment were similar (and are given in the Appendix), although in this case, the differences in performance between `GCM' and `DR-GCM' were less pronounced.

\subsection*{Experiment 3}

In a final experiment, we considered an implication of the previous asymptotic theoretical results that it may be preferable to estimate nuisance parameters inconsistently rather than at slow rates. We considered a data generating process closely related to one described in \citet{belloni2014inference}; $L$ was generated from a a multivariate normal distribution $\mathcal{N}(\mathbf{0},\Sigma)$, where $\Sigma_{k,j}=0.5^{|j-k|}$, $A\sim \mathcal{N}(\gamma^TL,1)$ and $Y\sim \mathcal{N}(\beta^TL,1)$, where $\gamma_j=\zeta_\gamma(1/j)^2$ and  $\beta_j=\zeta_\beta(1/j)^2$ for $j=1,...,200$. To estimate $m_0(L)$, we used the Lasso with penalties selected using the method proposed by \citet{belloni2014inference}, which we implemented using the `hdm' package in R \citep{chernozhukov2016hdm}. We considered four different approaches to estimate the propensity score; $i)$ using the Lasso, based on a correctly specified model; $ii)$  subtracting a random variable distributed as $\mathcal{N}(-3n^{-0.1},n^{-0.2})$ from the Lasso estimate; $iii)$ using the Lasso, after removing the first nine columns of $L$; $iv)$ as the mean of $A$. For the second approach, adding the random noise allowed us to slow down the rate of convergence of the propensity score estimator. The third and fourth approaches correspond to misspecifying the propensity score. We implemented the proposed `DR-GCM' test using each of these estimates of $g_0(L)$. The parameters $\zeta_\gamma$ and $\zeta_\beta$ were first both fixed at 0.82; we then considered a more challenging setting by lowering to $\zeta_\beta=0.2$, as here one might expect approaches that include a selection step for the exposure model to outperform those that omit such a step. We also reversed this, setting $\zeta_\beta=0.82$ and  $\zeta_\gamma=0.2$. The experiments were carried out at sample sizes  $n=$100, 250, 500, 1,000 and at each $n$, 1,000 simulated data sets were generated. 

\begin{figure}
  \caption{Third experiment, $\zeta_\gamma=\zeta_\beta=0.82$. \textcolor{black}{Black} dots represent standard Lasso estimation; circles represent $g_0$ estimated at a slow rate; triangles represent omission of the first nine covariates, crosses represent omission of all covariates.}
  \centering
    \includegraphics[width=0.55\textwidth]{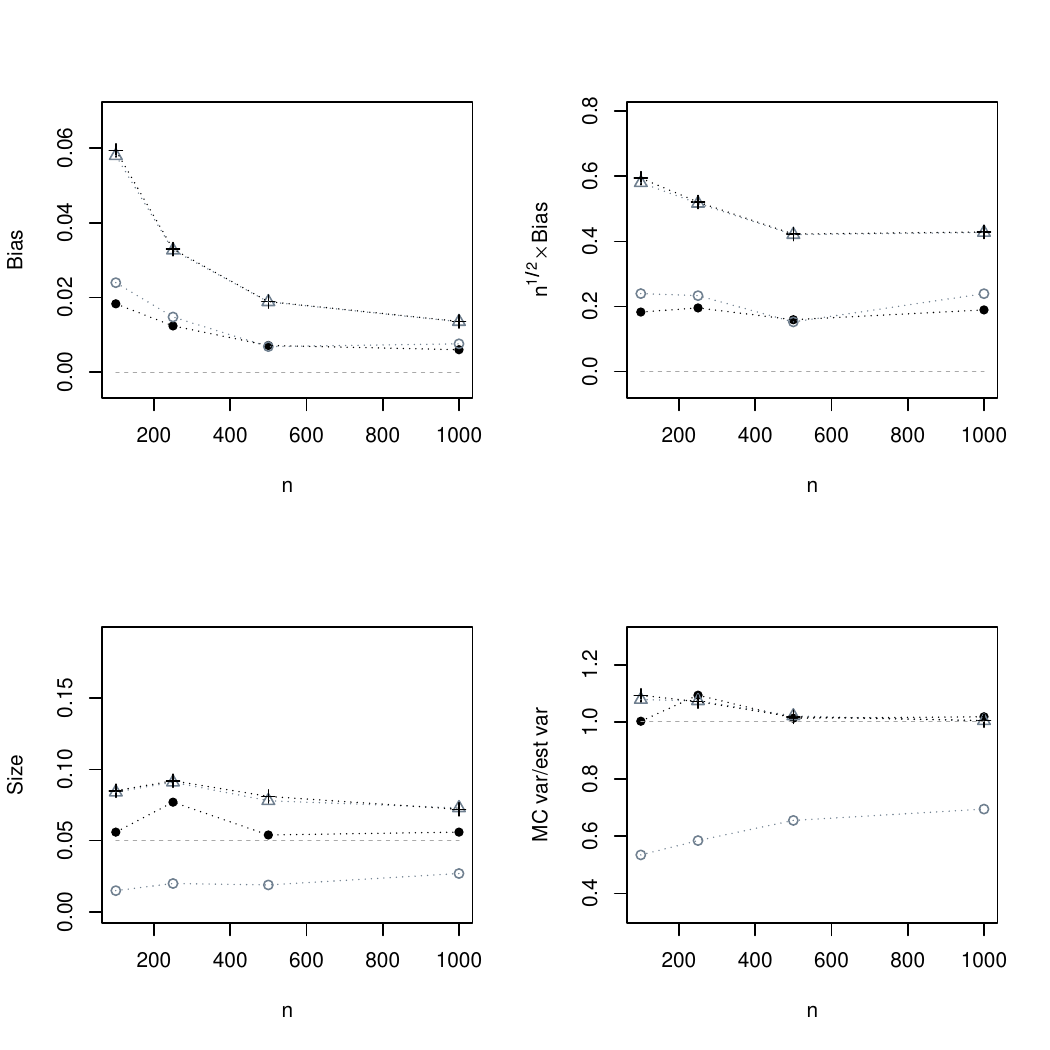}\label{revision_fit_a}
\end{figure}

\begin{figure}
  \caption{Third experiment,  $\zeta_\gamma=0.82$ $\zeta_\beta=0.2$. \textcolor{black}{Black} dots represent standard Lasso estimation; circles represent $g_0$ and $m_0$ estimated at a slow rate; triangles represent omission of the first nine covariates, crosses represent omission of all covariates.}
  \centering
    \includegraphics[width=0.55\textwidth]{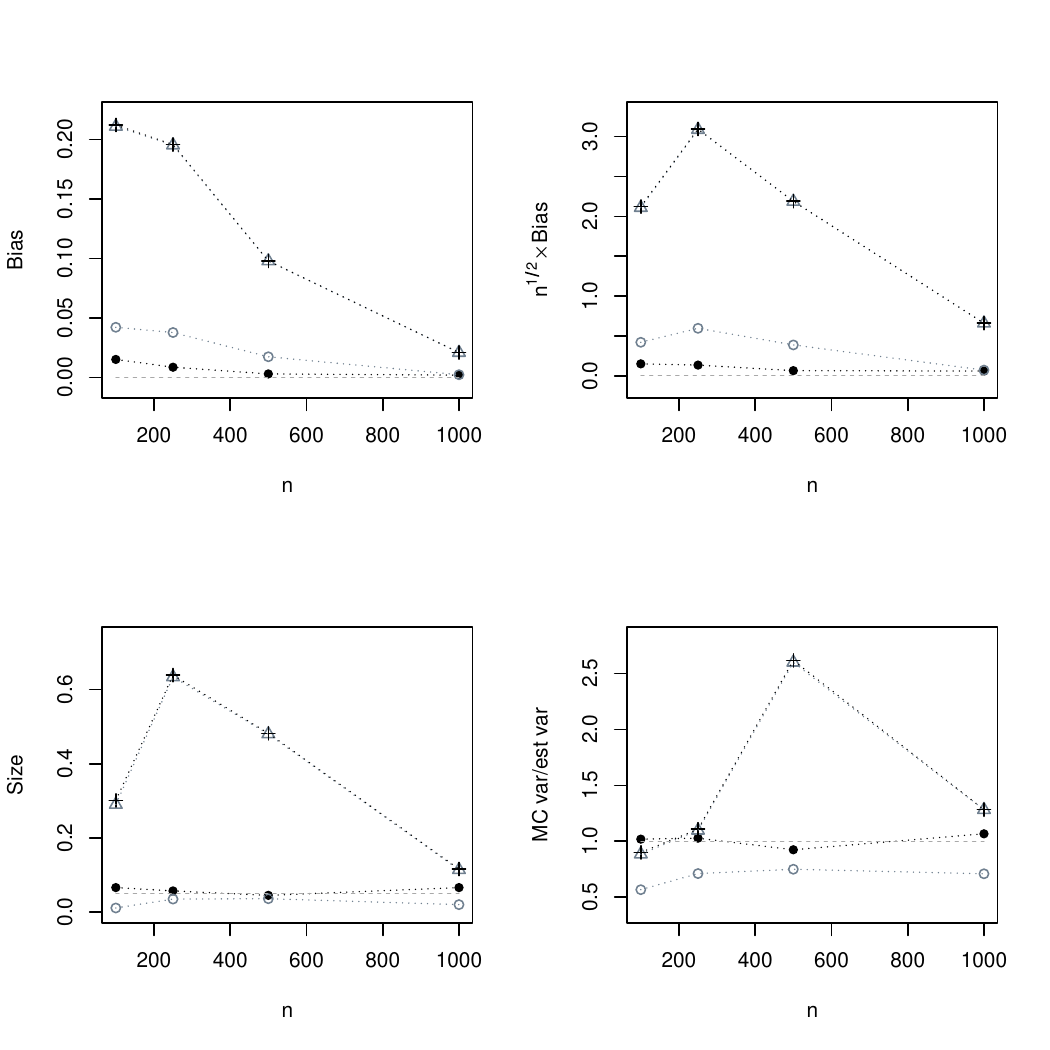}\label{revision_fit_b}
\end{figure}

\begin{figure}
  \caption{Third experiment,  $\zeta_\gamma=0.2$ $\zeta_\beta=0.82$. \textcolor{black}{Black} dots represent standard Lasso estimation; circles represent $g_0$, $m_0$ estimated at a slow rate; triangles represent omission of the first nine covariates, crosses represent omission of all covariates.}
  \centering
    \includegraphics[width=0.55\textwidth]{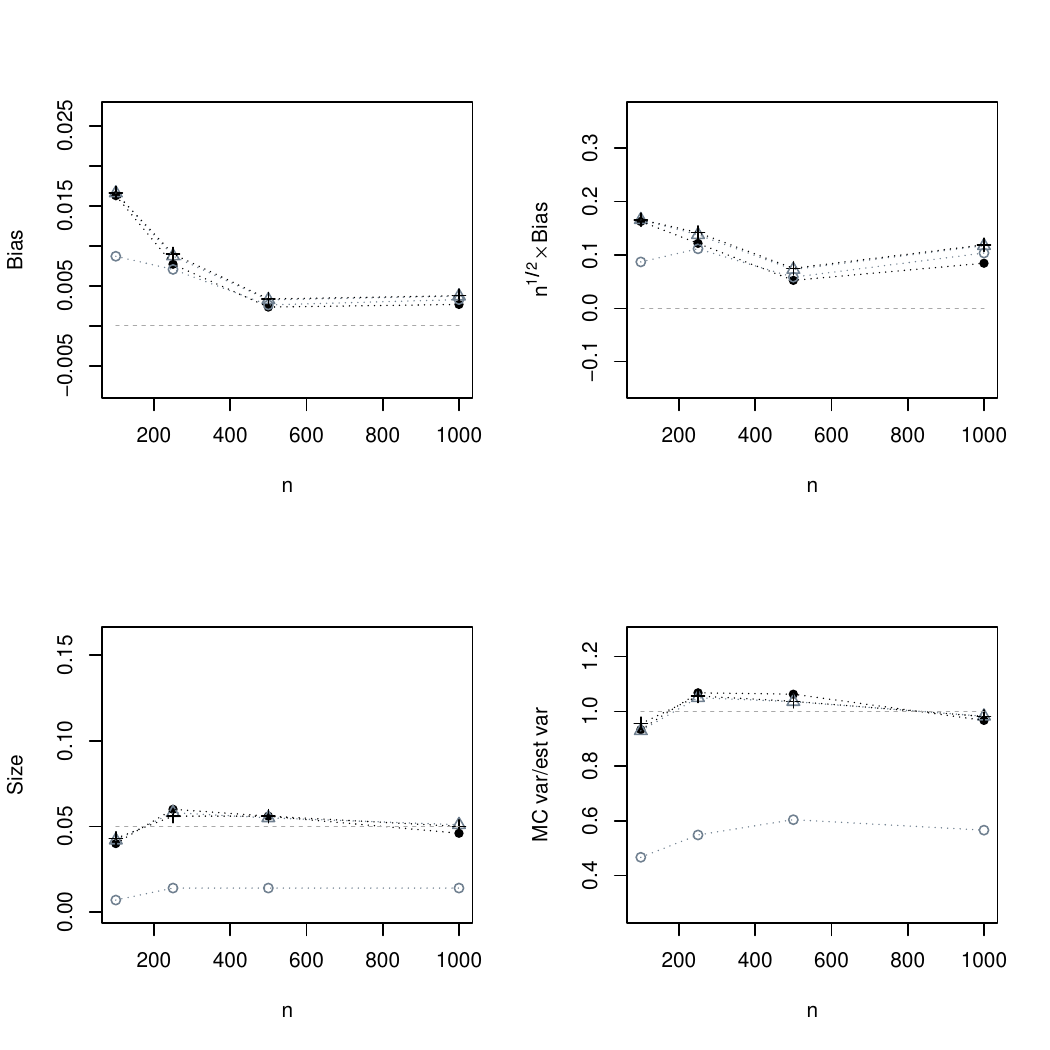}\label{revision_fit_c}
\end{figure}

The results are displayed in Figures \ref{revision_fit_a}, \ref{revision_fit_b} and \ref{revision_fit_c}. When $\zeta_\gamma=\zeta_\beta=0.82$, the approaches where the propensity score was misspecified had slightly inflated type 1 error at smaller sample sizes, due to a larger than desirable bias. However, performance improved as $n$ increased. If $\zeta_\gamma=0.2$, misspecification of the propensity score has negligible effect on the size of the test compared with fitting a correctly specified model. When the estimated propensity score converged to the truth at a slow rate, there was surprisingly low bias, but the estimated variance of the test statistic was generally an overestimate, leading to conservative performance. When $\zeta_\beta=0.2$ the differences in size were much starker between the different approaches, as misspecification was seen to lead to highly inflated type 1 error at low sample sizes, although this seemed to improve considerably at $n=1,000$. Hence although there may be settings where misspecification leads to a test attaining its size faster than using a slower converging estimator, we see that this is very sensitive to the data-generating process, and as a general statement does not reflect finite sample performance. This is compatible with the pointwise rather than uniform nature of the asymptotic results for doubly robust inference. 

Results in Section 4.2 indicate that it is possible to use the doubly robust inference framework to construct superefficient estimators. To assess this, in the setting where $\zeta_\gamma=\zeta_\beta=0.82$ and with $n=1,000$, we 
computed the influence function for an oracle semiparametric efficient estimator based on the true $g_0(L)$ and $m_0(L)$. We also computed the influence function for the doubly robust inference-based approach, and then estimated its variance over 1,000 simulations to assess how it compared to the semiparametric efficiency bound. In both cases, $\theta$ was fixed at the truth. When $g_0(L)$ was consistently estimated as in approach $(i)$ for estimating the propensity score, the ratio of the variance of the doubly robust inference influence function with the oracle influence function was 1.073. However, under approaches $(iii)$ and $(iv)$, the ratio of the variances changed to 0.269 and 0.268 respectively, providing clear evidence of superefficient behaviour.

\section{Discussion}

In this paper, we explored how standard doubly robust tests and estimators are asymptotically linear even when nuisance parameters are estimated using flexible data-adaptive methods, so long as they converge (sufficiently fast) to the truth. When at least one of the nuisance parameter estimators is inconsistent, the tests/confidence intervals of the target parameter will generally fail to possess their advertised size/coverage. Building on the work of \citet{benkeser2017doubly} and others, we have shown how to construct tests and estimators for semiparametric model parameters, which are asymptotically linear so long as at least one nuisance parameter is consistently estimated. The double robustness of the inferential procedures was borne out in simulation studies.

An important feature of the proposed estimators is that when only one nuisance is consistently estimated, they are only locally regular with respect to certain paths. We have characterised with respect to which submodels regularity is preserved. The sensitivity of these procedures to local changes in the data generating mechanism is perhaps not surprising. To give an intuitive example, suppose that $\hat{m}(L)$ is obtained using variable selection, $g^*(L)$ is deliberately misspecified and set to a constant, and the strength of association between a confounder and $Y$ (conditional on $A$ and other covariates) is of the order $n^{-1/2}$. In that case, standard selection procedures will eject the confounder in some but not all samples (at any finite $n$). Further, this confounder would not be picked up by the inconsistent estimator $\hat{g}(L)$, nor by $\hat{M}(L)$. As a result, the proposed estimator of $\theta_0$ may inherit the complex, non-standard behaviour of the $\hat{m}(L)$. For this reason, we also conjecture that when only one of $g_0(L)$ or $m_0(L)$ is consistently estimated,  our pointwise asymptotic results in Section 4 cannot be made uniform \citep{leeb2005model}. We therefore recommend using data-adaptive methods that have good performance across a range of function classes, and to aim for consistency for both nuisance estimators. 

Although the non-regularity might be viewed as a limitation of our proposal, we note that standard double machine learning/TMLE estimators are not generally regular and asymptotically linear when one nuisance is inconsistently estimated. In this case, our proposal is at least expected to deliver smaller bias and better coverage. An open question is whether data-adaptive doubly robust inference can be obtained without non-regularity; positive results exist for specific sparse estimators \citep{tan_model-assisted_2019,smucler_unifying_2019,dukes2020doubly,dukes2020inference,avagyan2021high}, but it is unclear whether they can be extended more generally. 

\begin{acks}
The authors thank Matteo Bonvini for helpful discussions and for alerting them to an error in Theorem 4 in a previous version of the paper. The first and second authors gratefully acknowledge support from BOF Grants BOF.01P08419 and BOF.24Y.2017.0004.01. The third author made his primary contributions to the paper prior to his current employment at GSK; he was previously employed at University College London where he was supported by the Wellcome trust [218554/Z/19/Z].
\end{acks}

\appendix

\section{Miscellaneous results}
 
\subsection{Expansion of the drift term in the partially logistic model}

Up to a scaling factor (and with abuse of notation), the drift term of the doubly robust estimator of \citet{tan2019doubly} is given by 
\begin{align*}
&P\left[\{1-\mathbb{E}(Y|L)\}\left(e^{m_0-\hat{m}}-1\right)(\nu_0-\hat{\nu})\right]\\
&=-P\left\{(1-Y)e^{m_0}\left(e^{-m_0}-e^{-\hat{m}}\right)(\nu_0-\hat{\nu})\right\}.
\end{align*}
By adding and subtracting some terms, the above equals
\begin{align*}
&P\left\{(1-Y)e^{m_0}\left(e^{-m^*}-e^{-m_0}\right)(\nu^*-\hat{\nu})\right)\\
&+P\left\{(1-Y)e^{m_0}\left(e^{-\hat{m}}-e^{-m^*}\right)(\nu_0-\nu^*)\right\}\\
&+P\left\{(1-Y)e^{m_0}\left(e^{-m^*}-e^{-m_0}\right)(\nu_0-\nu^*)\right\}\\
&+P\left\{(1-Y)e^{m_0}\left(e^{-m^*}-e^{-\hat{m}}\right)(\nu^*-\hat{\nu})\right\}\\
&=R_1+R_2+R_3+R_4.
\end{align*}
We will focus on terms $R_1$ and $R_2$. For the former, supposing that $\nu^*(L)=\nu_0(L)$, 
\begin{align*}
R_1&=P\left\{(1-Y)\left(e^{m_0-m^*}-1\right)(\nu_0-\hat{\nu})\right\}\\
&=P\left\{(Ye^{-\theta_0A-m^*}-(1-Y))(\nu_0-\hat{\nu})\right\}\\
&=P\left\{\mathbb{E}\{Ye^{-\theta_0A-m^*}-(1-Y)|\nu_0,\hat{\nu}\}(\nu_0-\hat{\nu})\right\}\\
&=P\left\{\mathbb{E}\{Ye^{-\theta_0A-m^*}-(1-Y)|\nu_0,\hat{\nu}\}(\nu_0-\hat{\nu})\frac{(1-Y)}{\{1-\mathbb{E}(Y|\nu_0,\hat{\nu})\}}\right\}\\
&=P\left\{\mathbb{E}\{Ye^{-\theta_0A-m^*}-(1-Y)|\nu_0,\hat{\nu}\}\{A-\hat{\nu}(L)\}\frac{(1-Y)}{\{1-\mathbb{E}(Y|\nu_0,\hat{\nu})\}}\right\}.
\end{align*}
Moving on to $R_2$, supposing that $m^*(L)=m_0(L)$,
\begin{align*}
R_2&=P\left\{(1-Y)\left(e^{m_0-\hat{m}}-1\right)(\nu_0-g^*)\right\}\\
&=P\left\{(1-Y)\left(e^{m_0-\hat{m}}-1\right)(A-g^*)\right\}\\
&=P\left\{(1-Y)\left(e^{m_0-\hat{m}}-1\right)\mathbb{E}(A-g^*|Y=0,m_0,\hat{m})\right\}\\
&=P\left[\{Ye^{-\theta_0A-\hat{m}}-(1-Y)\}\mathbb{E}(A-g^*|Y=0,m_0,\hat{m})\right].
\end{align*}

\subsection{Results for $\hat{\alpha}\hat{G}$}

\subsubsection{When $m^*(L)\neq m_0(L)$}

We will obtain results for the product $\hat{\alpha}_k\hat{G}_k$ in the case that $m^*(L)\neq m_0(L)$, since parallel developments can be shown straightforwardly for $\hat{\beta}_k\hat{M}_k$.  

We begin by considering the consistency of $\hat{\alpha}_k$.
Let $\phi(\alpha^*,g^*,G^*)=G^*(L)\{A-G^*(L)-\alpha^*G(L)\}$ and $\phi(\hat{\alpha}_k,\hat{g}^c_k,\hat{G}_k)$ denote the same quantity now indexed by estimates. Then
\begin{align*}
0&=\mathbb{P}_{n,k}\phi(\hat{\alpha}_k,\hat{g}^c_k,\hat{G}_k)-P\phi (\alpha^*,g^*,G^*)\\
&=(\mathbb{P}_{n,k}-P)\phi(\alpha^*,g^*,G^*)+
\mathbb{G}_{n,k}\{\phi(\hat{\alpha}_k,\hat{g}^c_k,\hat{G}_k)-\phi (\alpha^*,g^*,G^*)\}\\
&\quad+P\{\phi(\hat{\alpha}_k,\hat{g}^c_k,\hat{G}_k)-\phi (\alpha^*,g^*,G^*)\}\\
&=(\mathbb{P}_{n,k}-P)\phi(\alpha^*,g^*,G^*)+
\mathbb{G}_{n,k}\{\phi(\hat{\alpha}_k,\hat{g}^c_k,\hat{G}_k)-\phi (\alpha^*,g^*,G^*)\}\\
&\quad+P\{\phi(\hat{\alpha}_k,\hat{g}^c_k,\hat{G}_k)-\phi (\hat{\alpha}_k,g^*,G^*)\}+P\{\phi (\hat{\alpha}_k,g^*,G^*)-\phi (\alpha^*,g^*,G^*)\}.
\end{align*}
We will focus on the expression on the right hand side of the final equality. The first term $(\mathbb{P}_{n,k}-P)\phi(\alpha^*,g^*,G^*)$ is an i.i.d. sum of mean zero random variables, and therefore can be shown to be $O_P(n^{-1/2})$ under weak conditions. The second term $\mathbb{G}_{n,k}\{\phi(\hat{\alpha}_k,\hat{g}^c_k,\hat{G}_k)-\phi (\alpha^*,g^*,G^*)\}$ is an empirical process term; we conjecture that this can be handled in a similar way to the terms in the proof of Theorem \ref{key_theo}, and therefore expect it to be $o_P(n^{-1/2})$ given that sample splitting has been used. The final term can be shown to equal 
\[-(\hat{\alpha}_k-\alpha^*)PG^{*2}.\]

Then for the third term
\begin{align*}
&P\{\phi(\hat{\alpha}_k,\hat{g}^c_k,\hat{G}_k)-\phi (\hat{\alpha}_k,g^*,G^*)\}\\
&=P\hat{G}_k(g_0-\hat{g}^c_k)-PG^*(g_0-g^*)+P\hat{\alpha}_kP(G^{*2}-\hat{G}^2_k)\\
&=Pg^*(G^*-\hat{G}_k)+PG^*(g^*-\hat{g})+P(\hat{g}-g^*)(G^*-\hat{G}_k)\\&\quad +P\alpha^*P(G^{*2}-\hat{G}^2_k)+(\hat{\alpha}_k-\alpha^*)P(G^{*2}-\hat{G}^2_k)\\
&\lesssim \norm{\hat{G}_k-G^*}_{P,2}+\norm{\hat{g}^c_k-g^*}_{P,2}+(\hat{\alpha}_k-\alpha^*)\norm{\hat{G}_k-G^*}_{P,2}.
\end{align*} 

Hence by Slutsky's Theorem,
\begin{align*}
&(\mathbb{P}_{n,k}-P)\phi(\alpha^*,g^*,G^*)+
\mathbb{G}_{n,k}\{\phi(\hat{\alpha}_k,\hat{g}^c_k,\hat{G}_k)-\phi (\alpha^*,g^*,G^*)\}\\
&\quad+P\{\phi(\hat{\alpha}_k,\hat{g}^c_k,\hat{G}_k)-\phi (\hat{\alpha}_k,g^*,G^*)\}+P\{\phi (\hat{\alpha}_k,g^*,G^*)-\phi (\alpha^*,g^*,G^*)\}\\
&=(\mathbb{P}_{n,k}-P)\phi(\alpha^*,g^*,G^*)+o_P(n^{-1/2})+(\hat{\alpha}_k-\alpha^*)\left\{-PG^{*2}+O_P\left(\norm{\hat{G}_k-G^*}_{P,2}\right)\right\}\\
&\quad + O_P\left(\norm{\hat{G}_k-G^*}_{P,2}+\norm{\hat{g}^c_k-g^*}_{P,2}\right).
\end{align*}
Assuming the invertibility of 
\[-PG^{*2}+O_P\left(\norm{\hat{G}_k-G^*}_{P,2}\right),\]
we can rewrite the above as 
\begin{align*}
\hat{\alpha}_k-\alpha^*&=\left\{PG^{*2}+O_P\left(\norm{\hat{G}_k-G^*}_{P,2}\right)\right\}^{-1}\\
&\quad \times \bigg\{(\mathbb{P}_{n,k}-P)\phi(\alpha^*,g^*,G^*)+o_P(n^{-1/2})+ O_P\left(\norm{\hat{G}_k-G^*}_{P,2}+\norm{\hat{g}^c_k-g^*}_{P,2}\right)  \bigg\}.
\end{align*}
Hence, if $\hat{g}^c_k$ and $\hat{G}_k$ converge to a limit, then one would expect $\hat{\alpha}_k$ to also be consistent. Furthermore, the rate of $\hat{\alpha}_k$ will typically be determined by the slower of the rates of $\hat{g}^c_k$ and $\hat{G}_k$.

We note that Theorem \ref{key_theo} involves conditions on $\norm{\hat{\alpha}_k\hat{G}_k}_{P,2}$. Under the condition that $g^*(L)= g_0(L)$, which is required in this setting for asymptotical linearity of the test statistic, it follows that $\alpha^*=0$. In that case, 
\begin{align*}
\norm{\hat{\alpha}_k\hat{G}_k}_{P,2}=\hat{\alpha}_k\norm{\hat{G}_k}_{P,2}\leq&\hat{\alpha}_k\norm{\hat{G}_k-G^*}_{P,2}+\hat{\alpha}_k\norm{G^*}_{P,2}
\end{align*}
by the triangle inequality. The first term of the right hand side of the inequality will typically be of smaller order than the second, so $\norm{\hat{\alpha}_k\hat{G}_k}_{P,2}$ will typically be driven by the convergence rate of $\hat{\alpha}_k$.

\subsubsection{When $m^*(L)=m_0(L)$}

We will now consider the more challenging case where $m^*(L)=m_0(L)$ and $g^*(L)=g_0(L)$. We will now assume that the estimator $\hat{G}^c_k$ is obtained using the training sample, but expect this can be weakened along the lines of the proof of Theorem \ref{key_theo}. We give a result below that indicates that although the denominator in $\hat{\alpha}_k$ tends to zero, $|\mathbb{P}_{n,k}\hat{\alpha}_k\hat{G}^c_k|$ will in general converge to zero at a rate determined by the estimator $\hat{g}^c_k(L)$. 

We will make the following additional assumptions:
\begin{enumerate}[label=\roman*]
    \item Rates of convergence (in-sample error): \begin{align*}
    \norm{g_0-\hat{g}^c_k}_{\mathbb{P}_{n,k},2}&=O_P(n_k^{-a})\\
\norm{\hat{G}^c_k}_{\mathbb{P}_{n,k},2}
&=O_P(n_k^{-b})\\
\norm{\hat{G}^c_k}_{P,2}&=O_P(n_k^{-b})
\end{align*}
where $0<a \leq  1/2$ and $0<b \leq  1/2$. 
\item $\mathbb{P}_{n,k}\hat{G}^{c^2}_k$ can be inverted at any finite sample size.
\item $Var(A|L)<c<\infty $ with probability one where $c$ is a constant.
\end{enumerate}

Then with some abuse of notation,
\begin{align*}
|\mathbb{P}_{n,k}\hat{\alpha}_k\hat{G}^c_k|=&|\{\mathbb{P}_{n,k}\hat{G}^{c^2}_k\}^{-1}\mathbb{P}_{n,k}\{\hat{G}^c_k(A-\hat{g}^c_k)\}\mathbb{P}_{n,k}\hat{G}^c_k|\\
= &|\{\mathbb{P}_{n,k}\hat{G}^{c^2}_k\}^{-1}\mathbb{P}_{n,k}\{\hat{G}^c_k(A-g_0)\}\mathbb{P}_{n,k}\hat{G}^c_k+\{\mathbb{P}_{n,k}\hat{G}^{c^2}_k\}^{-1}\mathbb{P}_{n,k}\{\hat{G}_k(g_0-\hat{g}^c_k)\}\mathbb{P}_{n,k}\hat{G}^c_k|\\
\leq & |\{\mathbb{P}_{n,k}\hat{G}^{c^2}_k\}^{-1}\mathbb{P}_{n,k}\{\hat{G}^c_k(A-g_0)\}\mathbb{P}_{n,k}\hat{G}^c_k|+|\{\mathbb{P}_{n,k}\hat{G}^{c^2}_k\}^{-1}\mathbb{P}_{n,k}\{\hat{G}^c_k(g_0-\hat{g}^c_k)\}\mathbb{P}_{n,k}\hat{G}^c_k|.
\end{align*}
We will consider each term on the right hand side of the final equality in turn.

Firstly,
\begin{align*}
|\{\mathbb{P}_{n,k}\hat{G}^{c^2}_k\}^{-1}\mathbb{P}_{n,k}\{\hat{G}^c_k(A-g_0)\}\mathbb{P}_{n,k}\hat{G}^c_k|=&|\{\mathbb{P}_{n,k}\hat{G}^{c^2}_k\}^{-1}||\mathbb{P}_{n,k}\{\hat{G}^c_k(A-g_0)\}||\mathbb{P}_{n,k}\hat{G}^c_k|.
\end{align*}
By $(i)$, $\mathbb{P}_{n,k}\hat{G}^{c^2}_k=O_P(n_k^{-2b})$ and by $(ii)$, $\{\mathbb{P}_{n,k}\hat{G}^{c^2}_k\}^{-1}=O_P(n_k^{2b})$. Also, by $(i)$ and application of the Cauchy-Schwarz inequality, $|\mathbb{P}_{n,k}\hat{G}^c_k|=O_P(n_k^{-b})$. Moreover, 
\[\mathbb{E}\left[\mathbb{P}_{n,k}\{\hat{G}^c_k(A-g_0)\}|I^c_k\right]=0\] and by $(iii)$, 
\begin{align*}
Var\left[\mathbb{P}_{n,k}\{\hat{G}^c_k(A-g_0)\}|I^c_k\right]&=\frac{1}{n_k^2}\sum^{n_k}_{i=1} Var[\hat{G}_k^c(L_i)\{A_i-g_0(L_i)\}|I^c_k]\\
&=\frac{1}{n_k^2}\sum^{n_k}_{i=1} \mathbb{E}\{\hat{G}_k^c(L_i)^2Var(A_i|L_i)|I^c_k\}\\
&\leq c n_k^{-1} \norm{\hat{G}_k^c}_{P,2}^2=O_P(n_k^{-1-2b}).
\end{align*}
By application of Chebyshev's inequality, we have that 
\[|\mathbb{P}_{n,k}\{\hat{G}^c_k(A-g_0)\}|=O_P(n_k^{-1/2-b})=o_P(n_k^{-1/2})\]
Hence, we have
\[|\{\mathbb{P}_{n,k}\hat{G}^{c^2}_k\}^{-1}\mathbb{P}_{n,k}\{\hat{G}^c_k(A-g_0)\}\mathbb{P}_{n,k}\hat{G}^c_k|=O_P(n_k^{2b})O_P(n_k^{-1/2-b})O_P(n_k^{-b})=O_P(n_k^{-1/2}).\]

We now consider the term \[|\{\mathbb{P}_{n,k}\hat{G}^{c^2}_k\}^{-1}\mathbb{P}_{n,k}\{\hat{G}^c_k(g_0-\hat{g}^c_k)\}\mathbb{P}_{n,k}\hat{G}^c_k|=|\{\mathbb{P}_{n,k}\hat{G}^{c^2}_k\}^{-1}||\mathbb{P}_{n,k}\{\hat{G}^c_k(g_0-\hat{g}^c_k)\}||\mathbb{P}_{n,k}\hat{G}^c_k|.\] By the Cauchy-Schwarz inequality, we have that 
\[|\mathbb{P}_{n,k}\{\hat{G}^c_k(g_0-\hat{g}^c_k)\}|\leq \norm{\hat{G}^c_k}_{\mathbb{P}_{n,k},2}\norm{g_0-\hat{g}^c_k}_{\mathbb{P}_{n,k},2}= O_P(n_k^{-a-b}).\]
Hence, 
\[|\{\mathbb{P}_{n,k}\hat{G}^{c^2}_k\}^{-1}\mathbb{P}_{n,k}\{\hat{G}^c_k(g_0-\hat{g}^c_k)\}\mathbb{P}_{n,k}\hat{G}^c_k|=O_P(n_k^{2b})O_P(n_k^{-a-b})O_P(n_k^{-b})=O_P(n_k^{-a})\]
and by $(i)$,
\[|\mathbb{P}_{n,k}\hat{\alpha}_k\hat{G}^c_k|=O_P(n_k^{-a})+O_P(n_k^{-1/2})=O_P(n_k^{-a}).\]
It follows that this term inherits the rate of $\hat{g}^c_k$; although this will generally not be sufficiently fast to ensure that $\mathbb{P}_{n,k}\hat{\alpha}_k\hat{G}^c_k=o_P(n^{-1/2})$, we note that this is not necessarily problematic as $\mathbb{P}_{n,k}\hat{\alpha}_k\hat{G}^c_k$ appears in our expansion of the test statistic as a component of product terms alongside other quantities that also converge to zero.

\section{Proof of Theorem \ref{key_theo}}

\begin{proof}\textit{Proof of Theorem \ref{key_theo}}

We have
\begin{align*}
&\frac{1}{\sqrt{n}_k}\sum^{n_k}_{i=1}\{A_i-\hat{g}^c_k(L_i)-\hat{\alpha}_k\hat{G}_k(L_i)\}\{Y_i-\theta_0 A_i-\hat{m}^c_k(L_i)-\hat{\beta}_k \hat{M}_k(L_i)\}\\&\quad-\frac{1}{\sqrt{n}_k}\sum^{n_k}_{i=1}\{A_i-g^*(L_i)-\alpha^*G^*(L_i)\}\{Y_i-\theta_0 A_i-m^*(L_i)-\beta^* M^*(L_i)\}\\
&=\sqrt{n_k}\mathbb{P}_{n,k} \psi_1(\theta_0,\hat{\eta}^c_k,\hat{\tau}_k)-\sqrt{n_k}\mathbb{P}_{n,k} \psi_1(\theta_0,\eta^*,\tau^*)\\
&=\mathbb{G}_{n,k}\{\psi_1(\theta_0,\hat{\eta}^c_k,\hat{\tau}_k)-\psi_1(\theta_0,\eta^*,\tau^*)\}\\
&\quad+\sqrt{n_k}P \left\{\psi_1(\theta_0,\hat{\eta}^c_k,\hat{\tau}_k)-\psi_1(\theta_0,\eta^*,\tau^*)\right\}\\
&=\mathcal{I}_1+\mathcal{I}_2.
\end{align*} 
We will consider $\mathcal{I}_1$ and $\mathcal{I}_2$ in turn. 

\subsubsection*{Step 1}

We will proceed by showing that $Pr\left( |\mathbb{G}_{n,k}\left[\psi_1(\theta_0,\hat{\eta}^c_k,\hat{\tau}_k)-\psi_1(\theta_0,\eta^*,\tau^*)\right]|>\kappa \right)\to 0$ for any $\kappa > 0$ and therefore that $\mathcal{I}_1=o_P(1)$. Let us consider the class of functions \begin{align*}
\mathcal{F}&=[\psi_1\{\cdot;\theta_0,\hat{\eta}^c_k,\tau\}-\psi_1(\cdot;\theta_0,\eta^*,\tau^*):\tau \in \mathcal{T}]
\end{align*}
as well as the subclass
\[\mathcal{F}^{\delta_{n_k}}=[\psi_1\{\cdot;\theta_0,\hat{\eta}^c_k,\tau\}-\psi_1(\cdot;\theta_0,\eta^*,\tau^*):d_2(\tau,\tau^*)<\delta_{n_k}]\]
where $\delta_{n_k}\to 0$. 
Let $F_{\eta}$ be a measurable envelope for $f$ such that $\norm{F_\eta}_{P,q}< \infty$ for some $q\geq 2$, and $Z_{n_k}=\max_{i\leq n_k} F(W_i)$. Similarly, $F^{\delta_{n_k}}=F_{\hat{\eta}^c_k}+F_{\eta^*}$. 
Then for any $\kappa>0$,
\begin{align*}
&Pr\left[ |\mathbb{G}_{n,k}\left\{\psi_1(\theta_0,\hat{\eta}^c_k,\hat{\tau}_k)-\psi_1(W_i;\theta_0,\eta,\tau^*)\right\}|>\kappa \right]\\
&=\mathbb{E} \left(Pr\left[ |\mathbb{G}_{n,k}\left\{\psi_1(\theta_0,\hat{\eta}^c_k,\hat{\tau}_k)-\psi_1(W_i;\theta_0,\eta,\tau^*)\right\}|>\kappa | I^c_k \right]\right)\\
&=\mathbb{E} \left(Pr\left[ |\mathbb{G}_{n,k}\left\{\psi_1(\theta_0,\hat{\eta}^c_k,\hat{\tau}_k)-\psi_1(W_i;\theta_0,\eta,\tau^*)\right\}|>\kappa,d_2(\hat{\tau},\tau^*)<\delta_{n_k}| I^c_k \right]\right)\\
&\quad+\mathbb{E} \left(Pr\left[ |\mathbb{G}_{n,k}\left\{\psi_1(\theta_0,\hat{\eta}^c_k,\hat{\tau}_k)-\psi_1(W_i;\theta_0,\eta,\tau^*)\right\}|>\kappa,d_2(\hat{\tau},\tau^*)\geq\delta_{n_k}| I^c_k \right]\right)\\
&\leq \mathbb{E} \left\{Pr\left(\sup_{f \in \mathcal{F}^{\delta_{n_k}}}|\mathbb{G}_{n,k}(f)|>\kappa | I^c_k \right)\right\}+ P\left\{d_2(\hat{\tau},\tau^*)\geq \delta_{n_k}\right\}.
\end{align*}
By Assumption \ref{C_ML}, $Pr\left\{d_2(\hat{\tau},\tau^*)\geq \delta_{n_k}\ \right\}\to0$. Then it remains to show that the first term on the right hand side of the inequality goes to zero. 

From Corollary 5.1 of \citet{chernozhukov2014gaussian}, under Assumption \ref{EP} we have the bound on the uniform entropy integral
\[J(\delta_{n_k},\mathcal{F}^{\delta_{n_k}},F^{\delta_{n_k}})\leq 2\sqrt{2\nu}\delta_{n_k}\sqrt{\log(\xi/\delta_{n_k}) }\]
and the maximal inequality for VC-type classes
\[\mathbb{E}\left(\sup_{f \in \mathcal{F}^{\delta_{n_k}}}|\mathbb{G}_{n,k}(f)|\bigg|I^c_k \right)\lesssim  
 Cr^{(1)}_{n_k} \sqrt{\nu \log \left(\frac{\xi\norm{F^{\delta_{n_k}}}_{P,2}}{Cr^{(1)}_{n_k}}\right)}+\frac{\nu\norm{Z_{n_k}}_2}{\sqrt{n_k}} \log \left(\frac{\xi\norm{F^{\delta_{n_k}}}_{P,2}}{Cr^{(1)}_{n_k}}\right)\]
if $\sigma=Cr^{(1)}_{n_k}$ (for a constant $C$), where we use that $\sup_{f \in \mathcal{F}^{\delta_{n_k}}}\norm{f}_{P,2}=O_P(r^{(1)}_{n_k})$.  
By Assumption \ref{EP}, one can then convert this result into the asymptotic bound
\[\mathbb{E}\left(\sup_{f \in \mathcal{F}^{\delta_{n_k}}}|\mathbb{G}_{n,k}(f)|\bigg|I^c_k \right)=O_P\left(
r^{(1)}_{n_k} \sqrt{\log \left(\frac{1}{r^{(1)}_{n_k}}\right)}+\frac{1}{\sqrt{n_k}} \log \left(\frac{1}{r^{(1)}_{n_k}}\right)\right)=o_P(1).\]
Then by Markov's inequality, 
\begin{align*}
\mathbb{E} \left\{Pr\left(\sup_{f \in \mathcal{F}^{\delta_{n_k}}}|\mathbb{G}_{n,k}(f)|>\kappa \bigg| I^c_k \right)\right\}\leq \frac{1}{\kappa}\mathbb{E} \left\{\mathbb{E}\left(\sup_{f \in \mathcal{F}^{\delta_{n_k}}}|\mathbb{G}_{n,k}(f)|\bigg|I^c_k \right)\right\}=o(1)
\end{align*}
and it follows that $\mathcal{I}_1=o_P(1)$. 

\subsubsection*{Step 2}

One can show that 
\begin{align*}
\mathcal{I}_2&=\sqrt{n_k}P\left[\psi_1(\theta_0,\hat{\eta}^c_k,\hat{\tau}_k)-\psi_1(\theta_0,\eta^*,\tau^*)\right]\\
&=\sqrt{n_k}P\left\{\left(g_0-\hat{g}^c_k+\alpha^*G^*-\hat{\alpha}_k\hat{G}_k\right)\left(m_0-\hat{m}^c_k+\beta^* M^*-\hat{\beta}_k\hat{M}_k\right)\right\}.
\end{align*}
We will evaluate the above bias term in three settings:
\begin{enumerate}[label=(\roman*)]
\item $g_0(L)=g^*(L)$ and  $m_0(L)=m^*(L)$.
\item $g_0(L)\neq g^*(L)$ and  $m_0(L)=m^*(L)$.
\item $g_0(L)=g^*(L)$ and  $m_0(L)\neq m^*(L)$.
\end{enumerate}
In setting (i), by the Cauchy-Schwarz inequality, 
$\mathcal{I}_2=o_P(1)$ following Assumption \ref{RC_ML}. 

For setting (ii), we note first that $G^*(L)=0$ and $\beta^*=0$ since $m_0(L)=m^*(L)$. Then using Assumption \ref{RC_ML}, after splitting $\mathcal{I}_2$ as 
\begin{align*}
&\sqrt{n_k}P\left\{\left(g_0-\hat{g}^c_k\right)\left(m_0-\hat{m}^c_k-\hat{\beta}_k\hat{M}_k\right)\right\}\\
&-\sqrt{n_k}P\left\{\hat{\alpha}_k\hat{G}_k\left(m_0-\hat{m}^c_k-\hat{\beta}_k\hat{M}_k\right)\right\}
\end{align*}
we have that 
\begin{align*}
|\sqrt{n_k}P\left\{\hat{\alpha}_k\hat{G}_k\left(m_0-\hat{m}^c_k-\hat{\beta}_k\hat{M}_k\right)\right\}|=o_P(1).
\end{align*}
Also,
\begin{align*}
&\sqrt{n_k}P\left\{\left(g_0-\hat{g}^c_k\right)\left(m_0-\hat{m}^c_k-\hat{\beta}_k\hat{M}_k\right)\right\}\\
&=\sqrt{n_k}P\left\{\left(g_0-g^*\right)\left(m_0-\hat{m}^c_k-\hat{\beta}_k\hat{M}_k\right)\right\}\\
&\quad+\sqrt{n_k}P\left\{\left(g^*-\hat{g}^c_k\right)\left(m_0-\hat{m}^c_k-\hat{\beta}_k\hat{M}_k\right)\right\};
\end{align*}
the second term on the right hand side is $o_P(1)$. For the first term, with some abuse of notation,
\begin{align*}
&\sqrt{n_k}P\left\{\left(g_0-g^*\right)\left(m_0-\hat{m}^c_k-\hat{\beta}_k\hat{M}_k\right)\right\}\\
&=\sqrt{n_k}P\left\{\left(g_0-g^*\right)\left(m_0-\hat{m}^c_k-\hat{\beta}_k\hat{M}_k\right)\right\}\\
&=\sqrt{n_k}P\left\{(A-g^*)\left(m_0-\hat{m}^c_k-\hat{\beta}_k\hat{M}_k\right)\right\}\\
&=\sqrt{n_k}P\left\{\bar{M}\left(m_0-\hat{m}^c_k-\hat{\beta}_k\hat{M}_k\right)\right\}.
\end{align*}

After some algebraic manipulation,
\begin{align*}
&\sqrt{n_k}P\left\{\bar{M}\left(m_0-\hat{m}^c_k-\hat{\beta}_k\hat{M}_k\right)\right\}\\
&=\sqrt{n_k}\mathbb{P}_{n,k}\psi_2(\theta_0,\hat{\eta}^c_k,\hat{\tau}_k)-\sqrt{n_k}\mathbb{P}_{n,k}\psi_2(\theta_0,\eta^*,\tau^*)\\
&\quad-\mathbb{G}_{n,k}\left\{\psi_2(\theta_0,\hat{\eta}^c_k,\hat{\tau}_k)-\psi_2(\theta_0,\eta^*,\tau^*)\right\}\\
&\quad+\sqrt{n_k}P\{(M^*-\hat{M}_k)(m^*-\hat{m}^c_k-\hat{\beta}_k\hat{M}_k)\}+\sqrt{n_k}P\{(\bar{M}-M^*)(m^*-\hat{m}^c_k-\hat{\beta}_k\hat{M}_k)\}.
\end{align*}
The first term on the right hand side is exactly zero by virtue of how $\hat{\beta}$ is estimated. The two terms on the final row are $o_P(1)$ following Assumption \ref{RC_ML}; we again use the fact that $\beta M^*(L)$ is zero in this case. To handle the empirical process term, we will use similar arguments to in Step 1 of this proof; specifically, invoking Assumption \ref{EP} to show that 
$P\left[ |\mathbb{G}_{n,k}\left\{\psi_2(\theta_0,\hat{\eta}^c_k,\hat{\tau}_k)-\psi_2(\theta_0,\eta^*,\tau^*)\right\}|>\kappa \right]\to 0$. We apply the results in \citet{chernozhukov2014gaussian} now to the function class $\mathcal{F}_2=[\psi_j\{\cdot;\theta_0,\eta,\tau\}:\tau \in \mathcal{T}]$ for $m^*(L)=\hat{m}(L;I^c_k)$ and $m^*(L)=m^*(L)$ whilst conditioning on $I^c_k$. Combined with Assumption \ref{EP}, this indicates that 
\begin{align*}
&\sqrt{n_k}\mathbb{G}_{n,k}\left\{\psi_2(\theta_0,\hat{\eta}^c_k,\hat{\tau}_k)-\psi_2(\theta_0,\eta^*,\tau^*)\right\}=o_P(1).
\end{align*} 
Then we have shown that 
\begin{align*}
\sqrt{n_k}\mathbb{P}_{n,k} \left[\psi_1(\theta_0,\hat{\eta}^c_k,\hat{\tau}_k)\right]&=\sqrt{n_k}\mathbb{P}_{n,k} \left\{\psi_1(\theta_0,\eta^*,\tau^*)-\psi_2\{W_i;\theta_0,\eta^*,\tau^*\}\right\}+o_P(1)
\end{align*}
To finish the proof, one can repeat the same arguments for setting (iii) as for term (ii), in order to show that
\begin{align*}
\sqrt{n_k}\mathbb{P}_{n,k} \left\{\psi_1(\theta_0,\hat{\eta}^c_k,\hat{\tau}_k)\right\}&=\sqrt{n_k}\mathbb{P}_{n,k} \left\{\psi_1(\theta_0,\eta^*,\tau^*)\right\}\\&\quad-\sqrt{n_k}\mathbb{P}_{n,k}\left\{\psi_3(\theta_0,\eta^*,\tau^*)\right\}+o_P(1)
\end{align*}
in that setting.
\end{proof}

We are now in a position to prove Corollary \ref{key_cor}

\begin{proof}\textit{Proof of Corollary \ref{key_cor}}\\
Suppose for the moment that $g_0(L)\neq g^*(L)$. 
By the central limit theorem, 
\[\sqrt{n}_k \mathbb{P}_{n,k}\{\psi_1(\theta_0,\eta^*,\tau^*)-\psi_2(\theta_0,\eta^*,\tau^*)\}\]
converges to a normal distribution. 
Then following Theorem \ref{key_theo} and by application of Slutsky's theorem, it follows that 
$\sqrt{n_k}\mathbb{P}_{n,k} \psi_1(\theta_0,\hat{\eta}^c_k,\hat{\tau}_k)$ also converges to a normal distribution. Asymptotic normality can also be shown in the case that $m_0(L)\neq m^*(L)$ under similar reasoning.

It remains to show that 
\[\left[\mathbb{P}_{n,k}\psi^*(\theta_0,\hat{\eta}^c_k,\hat{\tau}_k)^2-\{\mathbb{P}_{n,k}\psi^*(\theta_0,\hat{\eta}^c_k,\hat{\tau}_k)\} ^2\right]^{-1}=\left\{P\psi^*(\theta_0,\eta^*,\tau^*)^2\right\}^{-1}+o_P(1)\]
Following Theorem \ref{key_theo}, we have that $\mathbb{P}_{n,k}\psi^*(\theta_0,\hat{\eta}^c_k,\hat{\tau}_k)=o_P(1)$ and by the continuous mapping theorem, $\{\mathbb{P}_{n,k}\psi^*(\theta_0,\hat{\eta}^c_k,\hat{\tau}_k)\}^2=o_P(1)$. 
Then we will show that 
\begin{align*}
\mathbb{P}_{n,k}\psi^*(\theta_0,\hat{\eta}^c_k,\hat{\tau}_k)^2-P\psi^*(\theta_0,\eta^*,\tau^*)^2=o_P(1)
\end{align*}
such that by application of the Slutsky’s theorem and the continuous mapping theorem, the main result follows.

By adding and subtracting terms,  
\begin{align*}
&\mathbb{P}_{n,k}\psi^*(\theta_0,\hat{\eta}^c_k,\hat{\tau}_k)^2-P\psi^*(\theta_0,\eta^*,\tau^*)^2\\&=\mathbb{P}_{n,k}\{\psi^*(\theta_0,\hat{\eta}^c_k,\hat{\tau}_k)^2-\psi^*(\theta_0,\eta^*,\tau^*)^2\}\\
&\quad+\mathbb{P}_{n,k}\psi^*(\theta_0,\eta^*,\tau^*)^2-P\psi^*(\theta_0,\eta^*,\tau^*)^2
\end{align*}
Then, 
\begin{align*}
&\mathbb{P}_{n,k}\{\psi^*(\theta_0,\hat{\eta}^c_k,\hat{\tau}_k)^2-\psi^*(\theta_0,\eta^*,\tau^*)^2\}\\
&=\mathbb{P}_{n,k} \left[\{\psi^*(\theta_0,\hat{\eta}^c_k,\hat{\tau}_k)-
\psi^*(\theta_0,\eta^*,\tau^*)\}\{\psi^*(\theta_0,\hat{\eta}^c_k,\hat{\tau}_k)+
\psi^*(\theta_0,\eta^*,\tau^*)\}\right]\\
&\leq  \left[\mathbb{P}_{n,k}\{\psi^*(\theta_0,\hat{\eta}^c_k,\hat{\tau}_k)-
\psi^*(\theta_0,\eta^*,\tau^*)\}^2\right]^{1/2} \left[\mathbb{P}_{n,k}\{\psi^*(\theta_0,\hat{\eta}^c_k,\hat{\tau}_k)+
\psi^*(\theta_0,\eta^*,\tau^*)\}^2\right]^{1/2}\\
&\leq  \norm{\psi^*(\theta_0,\hat{\eta}^c_k,\hat{\tau}_k)-
\psi^*(\theta_0,\eta^*,\tau^*)}_{\mathbb{P}_{n,k},2} \\
&\quad \times \bigg(\norm{\psi^*(\theta_0,\hat{\eta}^c_k,\hat{\tau}_k)-
\psi^*(\theta_0,\eta^*,\tau^*)}_{\mathbb{P}_{n,k},2}+2 \left[\mathbb{P}_{n,k}
\psi^*(\theta_0,\eta^*,\tau^*)^2\right]^{1/2}\bigg)
\end{align*}
Where we apply the Cauchy–Schwarz inequality and then the triangle inequality, using $(a+b)^2=\{(a-b)+2b\}^2$. 

We will next show that 
\begin{align*}
\norm{\psi^*(\theta_0,\hat{\eta}^c_k,\hat{\tau}_k)-
\psi^*(\theta_0,\eta^*,\tau^*)}_{\mathbb{P}_{n,k},2}=o_P(1).
\end{align*}
First, using the triangle inequality, 
\begin{align*}
&\norm{\psi^*(\theta_0,\hat{\eta}^c_k,\hat{\tau}_k)-
\psi^*(\theta_0,\eta^*,\tau^*)}_{\mathbb{P}_{n,k},2}\\
&\leq \norm{\psi^*(\theta_0,\hat{\eta}^c_k,\hat{\tau}_k)-
\psi^*(\theta_0,\hat{\eta}^c_k,\tau^*)}_{\mathbb{P}_{n,k},2}+\norm{\psi^*(\theta_0,\hat{\eta}^c_k,\tau^*)-
\psi^*(\theta_0,\eta^*,\tau^*)}_{\mathbb{P}_{n,k},2}
\end{align*}
For the first term on the right hand side of the equality, by Assumption \ref{EP}, it follows that this term is $o_P(1)$. For the second, 
\begin{align*}
\mathbb{E}\left\{ \norm{\psi^*(\theta_0,\hat{\eta}^c_k,\tau^*)-
\psi^*(\theta_0,\eta^*,\tau^*)}^2_{\mathbb{P}_{n,k},2}|I^c_k\right\}=\norm{\psi^*(\theta_0,\hat{\eta}^c_k,\tau^*)-
\psi^*(\theta_0,\eta^*,\tau^*)}^2_{P,2}.
\end{align*}
By Assumption \ref{C_ML}, this can be shown to be $o_P(1)$. Hence, by Markov's inequality, \[\norm{\psi^*(\theta_0,\hat{\eta}^c_k,\tau^*)-
\psi^*(\theta_0,\eta^*,\tau^*)}^2_{\mathbb{P}_{n,k},2}=o_P(1)\] and by the continuous mapping theorem, the same holds for $\norm{\psi^*(\theta_0,\hat{\eta}^c_k,\tau^*)-
\psi^*(\theta_0,\eta^*,\tau^*)}_{\mathbb{P}_{n,k},2}$. Then, the result for $\norm{\psi^*(\theta_0,\hat{\eta}^c_k,\hat{\tau}_k)-
\psi^*(\theta_0,\eta^*,\tau^*)}_{\mathbb{P}_{n,k},2}$ follows. 

Next, using the weak law of large numbers, 
\[\mathbb{P}_{n,k}\psi^*(\theta_0,\eta^*,\tau^*)^2-P\psi^*(\theta_0,\eta^*,\tau^*)^2=o_P(1)\]
so that $\mathbb{P}_{n,k}
\psi^*(\theta_0,\eta^*,\tau^*)^2=O_P(1)$. Hence,
\[\mathbb{P}_{n,k}\{\psi^*(\theta_0,\hat{\eta}^c_k,\hat{\tau}_k)^2-\psi^*(\theta_0,\eta^*,\tau^*)^2\}=o_P(1)\left\{o_P(1)+O_P(1)\right\}=o_P(1).\]
Given that we have already shown the result for $\mathbb{P}_{n,k}\psi^*(\theta_0,\eta^*,\tau^*)^2-P\psi^*(\theta_0,\eta^*,\tau^*)^2$, the desired result follows.

\end{proof}

\section{Proof of Theorem \ref{reg_theo}}
\begin{proof}
For any $s_\theta$, we have that $n^{1/2}(\hat{\theta}_n-\theta_n) = n^{1/2}(\hat{\theta}_n-\theta_0) - s_\theta$. It follows from local asymptotic linearity that under sampling from $P$ we have
$$\begin{pmatrix}
		n^{1/2}(\hat{\theta}_n-\theta_n) \\
		\Lambda_n
		\end{pmatrix} \overset{P^n}{\rightsquigarrow} N\left( \begin{pmatrix}
		-s_\theta \\ - \frac{1}{2} P(\mathbb{B}s)^2
		\end{pmatrix} ,
		\begin{pmatrix}
		\tau_*^2 & \rho_* \\ \rho_* & P(\mathbb{B}s)^2
	\end{pmatrix} \right)\,.$$
Here, $\rho_*$ is the asymptotic covariance between $\mathbb{B}s$ and the influence function of $\hat{\theta}_n$ and $\tau_*^2$ is variance of the same influence function. 

By Le Cam's third lemma, we have that under sampling from $P_{\theta_{n^{-1/2}},\eta_{n^{-1/2}}}$ that
$$ 	n^{1/2}(\hat{\theta}_n-\theta_n) \overset{}{\rightsquigarrow} N(\mu_s, \tau^2_*), $$
where $\mu_s = \rho_* - s_\theta$. There are three cases, corresponding to (1) both $m_0(L)$ and $g_0(L)$ consistently estimated, (2) just $m_0(L)$ and (3) just $g_0(L)$ consistently estimated.

Below, we let $\epsilon_0(W) = Y-A\theta_0-m_0(L)$, $\delta_0(W) = A -g_0(L)$, $\epsilon_*(W) = Y-A\theta_0-m^*(L)$, and $\delta_*(W) = A -g^*(L)$. Moreover, we define
\begin{align}
	u_0(W) &= \epsilon_0(W) \delta_0(W) \label{u0} \\ 
	u_{\epsilon}(W) &= \epsilon_0(W) \{\delta_*(W)-M^*(L)\} = u_0(W) + \{\delta_*(W)-\delta_0(W)-M^*(L)\}\epsilon_0(W) \label{ue} \\ 
	u_{\delta}(W) &= \{\epsilon_*(W)-G^*(L)\}\delta_0(W) = u_0(W) + \{\epsilon_*(W)-\epsilon_0(W)-G^*(L)\}\delta_0(W) \label{ud}
\end{align}
as shorthand for the influence function of the score statistic and let $\dot{u}$ denote the partial derivative of a function $u$ with respect to $\theta$.
		
\paragraph{Case (1).} We have that $\rho_* = -P_0 (u_0 \mathbb{B} s)/P\dot{u}_0$. The numerator equals (with abuse of notation)
\begin{align*}
	P (u_0 \mathbb{B} s) &= P [\epsilon_0 \delta_0\{\epsilon_0(A s_\theta+s_m)+\delta_0 s_g\}] \\
	&= P \epsilon_0^2 \delta_0 As_\theta + P \epsilon_0^2 \delta_0 s_m + P \epsilon_0 \delta_0^2 s_g \\
	&= P \delta_0 A s_\theta + P \delta_0 s_m + 0 \\
	&= s_\theta P \{\delta_0 A\}\,.
\end{align*}
The penultimate equality follows from $\mathbb{E}\{ \epsilon_0(O) \mid A, L\} = 0$ and $\mathbb{E}\{ \epsilon_0(O)^2 \mid A, L\} = 1$ for the present setting. The denominator is simply
\begin{align*}
	-P \dot{u}_0(W) &= -P \frac{\partial}{\partial \theta} (Y-A \theta-m_0)\delta_0 \\
	&= P\{\delta_0 A\}.
\end{align*}
Hence, $\rho_* = s_\theta$ in this case and $\mu_s = 0$ as claimed.
		
\paragraph{Case (2).} Here, we have that $\rho_* = P \{u_\epsilon(W) \mathbb{B} s(W)\}/P\dot{u}_\epsilon(W)$. The numerator equals (with abuse of notation)
\begin{align*}
	P (u_0 \mathbb{B}_0 s) &= P [ \epsilon_0(\delta_*-M^*)\{\epsilon_0(As_\theta+s_m)+\delta_0 s_g\}] \\
	&= P \epsilon_0^2(\delta_*-M^*)As_\theta + P \epsilon_0^2(\delta_*-M^*)s_m + 0  \\
	&= s_\theta P \{(\delta_*-M^*)A\} + P (\delta_*-M^*)s_m \,.
\end{align*}
For the denominator, we have
\begin{align*}
	-P \dot{u}_\epsilon &= -P \frac{\partial}{\partial \theta} (Y-A \theta-m_0)(\delta_*-M^*) \\
	&= P\{(\delta_*-M^*)A\}.
\end{align*}
Putting this together, we have $\rho_* = s_\theta + P\{(\delta_*-M^*)s_m\}/P\{(\delta_*-M^*)A\}$. The asymptotic mean is $\mu_s = P\{(\delta_*-M^*)s_m\}/P\{(\delta_*-M^*)A\}$.
		
\paragraph{Case (3).} Lastly, we have that $\rho_* = P (u_\delta \mathbb{B} s)/P\dot{u}_\delta$. The numerator equals (with abuse of notation)
\begin{align*}
	P (u_0 \mathbb{B} s) &= P [ \delta_0(\epsilon_*-G^*)\{\epsilon_0(As_\theta+s_m)+\delta_0 s_g\}] \\
	&= P \epsilon_0\delta_0(\epsilon_*-G^*)(As_\theta+s_m) + P \delta_0^2(\epsilon_*-G^*)s_g  \\
	&= s_\theta P \epsilon_*\epsilon_0 \delta_0 A + P \epsilon_*\epsilon_0 \delta_0 s_m + P(\epsilon_*-G^*)s_g \\
	&= s_\theta P (\delta_0 A) + 0 + P(\epsilon_*-G^*)s_g \,.
\end{align*}
Here, we have used that $\mathbb{E}\{ \epsilon_*(O)\epsilon_0(O) \mid A, L\} = 1$ in the present setting. For the denominator, we have
\begin{align*}
	-P \dot{u}_\epsilon(W) &= -P \frac{\partial}{\partial \theta} \{Y-A \theta-m^*-G^*\}\delta_0 \\
	&= P_0(\delta_0 A).
\end{align*}
Putting this together, we have $\rho_* = s_\theta + P\{(\epsilon_*-G^*)s_g\}/P(\delta_0A)$. The asymptotic mean is $\mu_s = P\{(\epsilon_*-G^*)s_g\}/P(\delta_0A)$.
		
This completes the proof.
\end{proof}

\section{Proof of Theorem \ref{kern_theo}}

\begin{proof}
Define the oracle estimator as
\begin{align*}
\tilde{M}(x)
&=\hat{f}^{-1}_{m,n_k}(x)\mathbb{P}_{n,k}\left[\varphi_2(x,h,\eta^*)\right].
\end{align*}
Our proof is based around the expansion
\begin{align*}
\hat{M}(x)-M^*(x)&=\hat{M}(x)-\tilde{M}(x)+\tilde{M}(x)-M^*(x).
\end{align*}
Using conditions \ref{C_K1}-\ref{C_CD}, it follows from standard theory on higher order kernel estimators (see e.g. Section 1.11 of \cite{li2023nonparametric}) that 
\[|\tilde{M}(x)-M^*(x)|=O_P\left(\frac{1}{\sqrt{n_kh}}+h^\vartheta\right).\]
Notably, the bias of $\tilde{M}(x)$ is $O(h^\vartheta)$ and the variance is $O(1/nh)$. We will now focus on the difference $\hat{M}(x)-\tilde{M}(x)$:
\begin{align*}
&\hat{M}(x)-\tilde{M}(x)\\=&\hat{f}^{-1}_{\hat{m},n_k}(x)P\left\{\varphi_2(x,h,\hat{\eta}^c_k)-\varphi_2(x,h,\eta^*)\right\}\\
&+\hat{f}^{-1}_{\hat{m},n_k}(x)(\mathbb{P}_{n,k}-P)\left\{\varphi_2(x,h,\hat{\eta}^c_k)-\varphi_2(x,h,\eta^*)\right\}\\
&+(\hat{f}^{-1}_{\hat{m},n_k}-\hat{f}^{-1}_{m,n_k})\mathbb{P}_{n,k}\left\{\varphi_2(x,h,\eta^*)\right\}
\end{align*}
In order to obtain (\ref{ktres1}), we will show in turn that
\begin{align}
P\left\{\varphi_2(x,h,\hat{\eta}^c_k)-\varphi_2(x,h,\eta^*)\right\}&=O_P\left(\zeta_g+h^{-1}\zeta_m\right)\label{k_res_1}\\
(\mathbb{P}_{n,k}-P)\left\{\varphi_2(x,h,\hat{\eta}^c_k)-\varphi_2(x,h,\eta^*)\right\}&=o_P\left(\frac{1}{\sqrt{n_kh}}\right)\label{k_res_2}\\
\hat{f}^{-1}_{\hat{m},n_k}-\hat{f}^{-1}_{m,n_k}&=O_P\left(h^{-1}\zeta_m\right)+o_P\left(\frac{1}{\sqrt{n_kh}}\right). \label{k_res_3}
\end{align}

Note that by the triangle inequality,
\begin{align*}
&|P\left\{\varphi_2(x,h,\hat{\eta}^c_k)-\varphi_2(x,h,\eta)\right\}|\\&=O_P\bigg(\bigg|\int \left\{K\left(\frac{x-\hat{m}^c_k(l)}{h}\right)-K\left(\frac{x-m^*(l)}{h}\right)\right\}\{g_0(l)-g^*(l)\}dP(w)\bigg|\bigg)\\
&\quad+O_P\bigg(\bigg|\int K\left(\frac{x-m^*(l)}{h}\right)\{g^*(l)-\hat{g}^c_k(l)\}dP(w)\bigg|\bigg)\\
&\quad+O_P\bigg(\bigg|\int \left\{K\left(\frac{x-\hat{m}^c_k(l)}{h}\right)-K\left(\frac{x-m^*(l)}{h}\right)\right\}\{g^*(l)-\hat{g}^c_k(l)\}dP(w)\bigg|\bigg).
\end{align*}
Then, by condition \ref{C_K2}, \ref{ms_b} and the Cauchy-Schwarz inequality, 
\begin{align*}&\bigg|\int \left\{K\left(\frac{x-\hat{m}^c_k(l)}{h}\right)-K\left(\frac{x-m^*(l)}{h}\right)\right\}\{g_0(l)-g^*(l)\}dP(w)\bigg|\\
&\leq \bigg|\int Ch^{-1}|m^*(l)-\hat{m}^c_k(l)| |g_0(l)-g^*(l)|dP(w)\bigg|\\
&\lesssim  h^{-1}\norm{m^*-\hat{m}^c_k}_{P,2}=O_P(h^{-1}\zeta_m)
\end{align*}
where $C$ is a constant. Using similar reasoning, one can show that 
\begin{align*}
O_P\bigg(\bigg|\int K\left(\frac{x-m^*(l)}{h}\right)\{g^*(l)-\hat{g}^c_k(l)\}dP(w)\bigg|\bigg)&=O_P(\zeta_g)\\
O_P\bigg(\bigg|\int \left\{K\left(\frac{x-\hat{m}^c_k(l)}{h}\right)-K\left(\frac{x-m^*(l)}{h}\right)\right\}\{g^*(l)-\hat{g}^c_k(l)\}dP(w)\bigg|\bigg)&=O_P(h^{-1}\zeta_g\zeta_m)
\end{align*}
and (\ref{k_res_1}) follows. 

Since $K$ is fixed, we also have that
\begin{align*}
&\sup_{x \in \mathcal{X}}|\sqrt{n_kh}(\mathbb{P}_{n,k}-P)\left[\varphi_2(x,h,\hat{\eta}^c_k)-\varphi_2(x,h,\eta^*)\right]|\\
&\lesssim \max_{k}\sup_{f\in\mathcal{F}_{n,k}}|\sqrt{h}\mathbb{G}_{n,k}(f)|
\end{align*}
where $\mathcal{F}_{n,k}=\left\{\varphi_2(\cdot;x,h,\hat{\eta}^c_k)-\varphi_2(\cdot;x,h,\eta^*):x\in\mathcal{X} \right\}$
with the corresponding envelope function 
$F_{n,k}=\max_{k}\sup_{x\in \mathcal{X}}|\left\{\varphi_2(L;x,h,\hat{\eta}^c_k)-\varphi_2(L;x,h,\eta^*)\right\}|$. 
By condition \ref{C_ent} on the kernel, the class $\{K(x-\tilde{m}(L))/h)\{A-\tilde{g}(L)\}: x\in \mathcal{X}\}$ is of VC-type for fixed $\tilde{m}(L)=m^*(L)$, $\tilde{m}(L)=\hat{m}(L;I^c_k)$, $\tilde{g}(L)=g^*(L)$ and $\tilde{g}(L)=\hat{g}(L;I^c_k)$. 
By the permanence properties of VC classes \citep{van1996weak}, it follows that $\mathcal{F}_{n,k}$ is also of VC-type and hence has a polynomial covering number. We can then invoke Corollary 5.1 of \citet{chernozhukov2014gaussian}, such that via conditioning on the auxiliary sample,
\[\mathbb{E}\left\{\sup_{f \in \mathcal{F}_{n,k}}|\sqrt{h}\mathbb{G}_{n,k}(f)|\bigg|I^c_k\right\}\lesssim
\sqrt{h}t^{(2)}_{n_k} \sqrt{\log \left(\frac{1}{\sqrt{h}t^{(2)}_{n_k} }\right)}+\frac{1}{\sqrt{n_k}} \log \left(\frac{1}{\sqrt{h}t^{(2)}_{n_k} }\right).\]
by condition \ref{C_con}. Using Markov's inequality (along the lines of the previous proof), one can therefore show that $\sqrt{h}\mathbb{G}_{n,k}(f)=o_P(1)$. Therefore we have (\ref{k_res_2}).

Also,
\begin{align*}
\hat{f}^{-1}_{\hat{m},n_k}-\hat{f}^{-1}_{m,n_k}&=\hat{f}^{-1}_{\hat{m},n_k}(\hat{f}_{m,n_k}-\hat{f}_{\hat{m},n_k})\hat{f}^{-1}_{\hat{m},n_k}.
\end{align*}
First, 
\begin{align*}
\hat{f}_{\hat{m},n_k}-\hat{f}_{m,n_k}=&P\left\{K\left(\frac{x-\hat{m}^c_k}{h}\right)-K\left(\frac{x-m^*}{h}\right)\right\}\\
&+(\mathbb{P}_{n,k}-P)\left\{K\left(\frac{x-\hat{m}^c_k}{h}\right)-K\left(\frac{x-m^*}{h}\right)\right\}
\end{align*}
Using the same arguments as for (\ref{k_res_1}),
\[P\left\{K\left(\frac{x-\hat{m}^c_k}{h}\right)-K\left(\frac{x-m^*}{h}\right)\right\}=O_P\left(h^{-1}\zeta_m\right)\]
by condition \ref{C_K2}. Furthermore, one can show that 
\[(\mathbb{P}_{n,k}-P)\left\{K\left(\frac{x-\hat{m}^c_k}{h}\right)-K\left(\frac{x-m^*}{h}\right)\right\}=o_P\left(\frac{1}{\sqrt{n_kh}}\right)\]
if conditions \ref{C_ent} and \ref{C_con} hold. 
Finally, a standard result on kernel density estimators is that
\[\mathbb{E}[\{\hat{f}_{m,n_k}(x)-f_{m,n_k}(x)\}^2]=O(h)+O\left(\frac{1}{n_kh}\right)\]
and by Assumptions \ref{C_K1}, it follows that $\mathbb{E}\{(\hat{f}_{m,n_k}-f_{m,n_k})^2\}=o(1)$, and thus $\hat{f}_{m,n_k}=O_P(1)$ and moreover $\hat{f}_{\hat{m},n_k}=O_P(1)$. Invoking condition \ref{C_K3}, we have shown that 
\begin{align*}
\hat{f}^{-1}_{\hat{m},n_k}-\hat{f}^{-1}_{m,n_k}&=O_P(1)\left\{O_P\left(h^{-1}\zeta_m\right)+o_P\left(\frac{1}{\sqrt{n_kh}}\right)\right\}O_P(1).
\end{align*}
Using the same reasoning, one can show that $\mathbb{P}_{n,k}\left[\varphi_2(x,h,\eta)\right]=O_P(1)$ and result (\ref{ktres1}) follows.

To show result (\ref{ktres2}), note that the previous results imply that 
\[\mathbb{E}\{\hat{M}(x)\}-M^*(x)=O(h^\vartheta+\zeta_g+h^{-1}\zeta_m)\]
and 
\[\mathbb{E}\left([\hat{M}(x)-\mathbb{E}\{\hat{M}(x)\}]^2\right)=O\left(\frac{1}{n_kh}\right).\]
\end{proof}

\section{Additional simulation results}

\begin{figure}[h]
  \caption{Second experiment (all nuisances estimated consistently). \textcolor{black}{Black} dots represent `PS'; \textcolor{red}{red} squares represent `OR'; \textcolor{green}{green} triangles represent `DML'; \textcolor{blue}{blue} triangles with represent `DR-DML'.}
  \centering
    \includegraphics[width=0.55\textwidth]{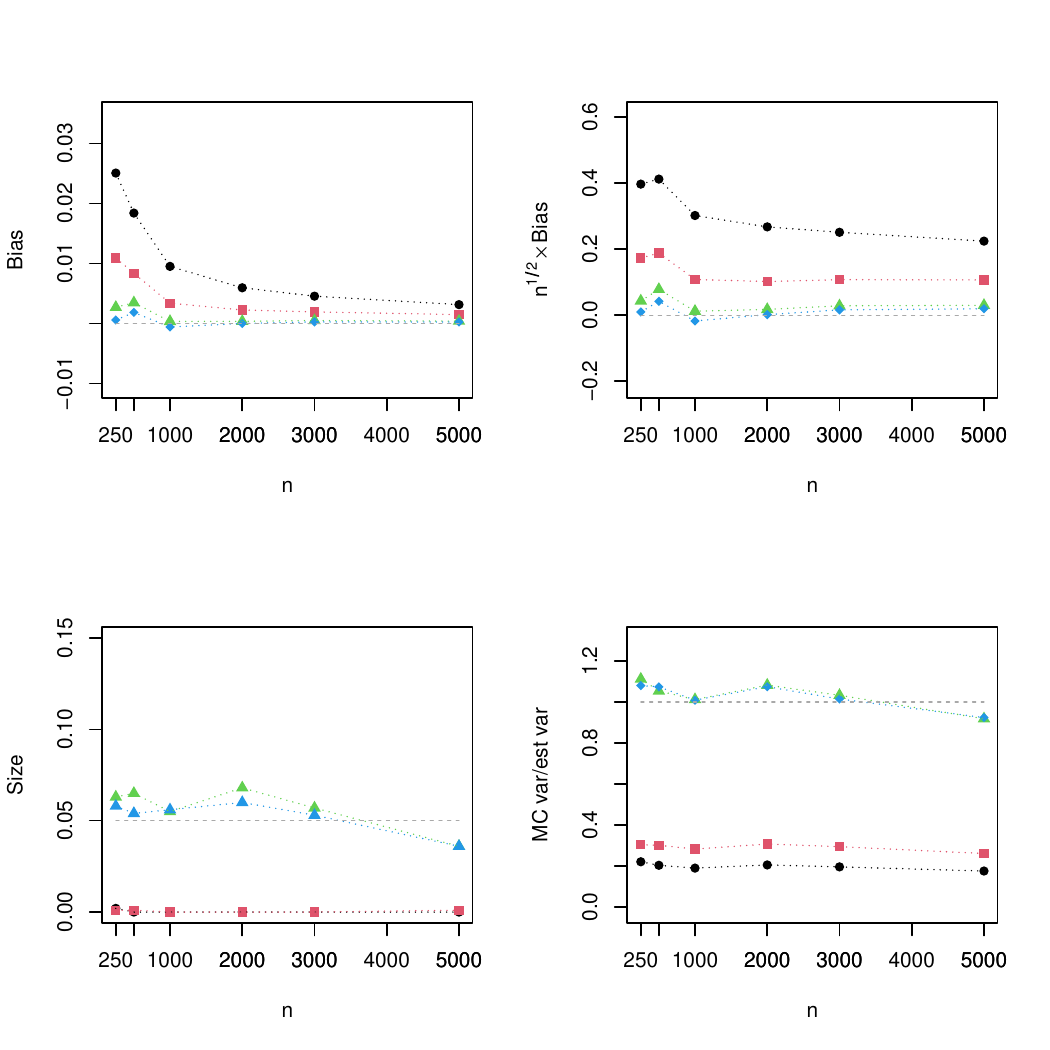}
\end{figure}

\begin{figure}
  \caption{Second experiment (propensity score estimated correctly). \textcolor{black}{Black} dots represent `PS'; \textcolor{green}{green} triangles represent `DML'; \textcolor{blue}{blue} triangles with represent `DR-DML'.}
  \centering
    \includegraphics[width=0.55\textwidth]{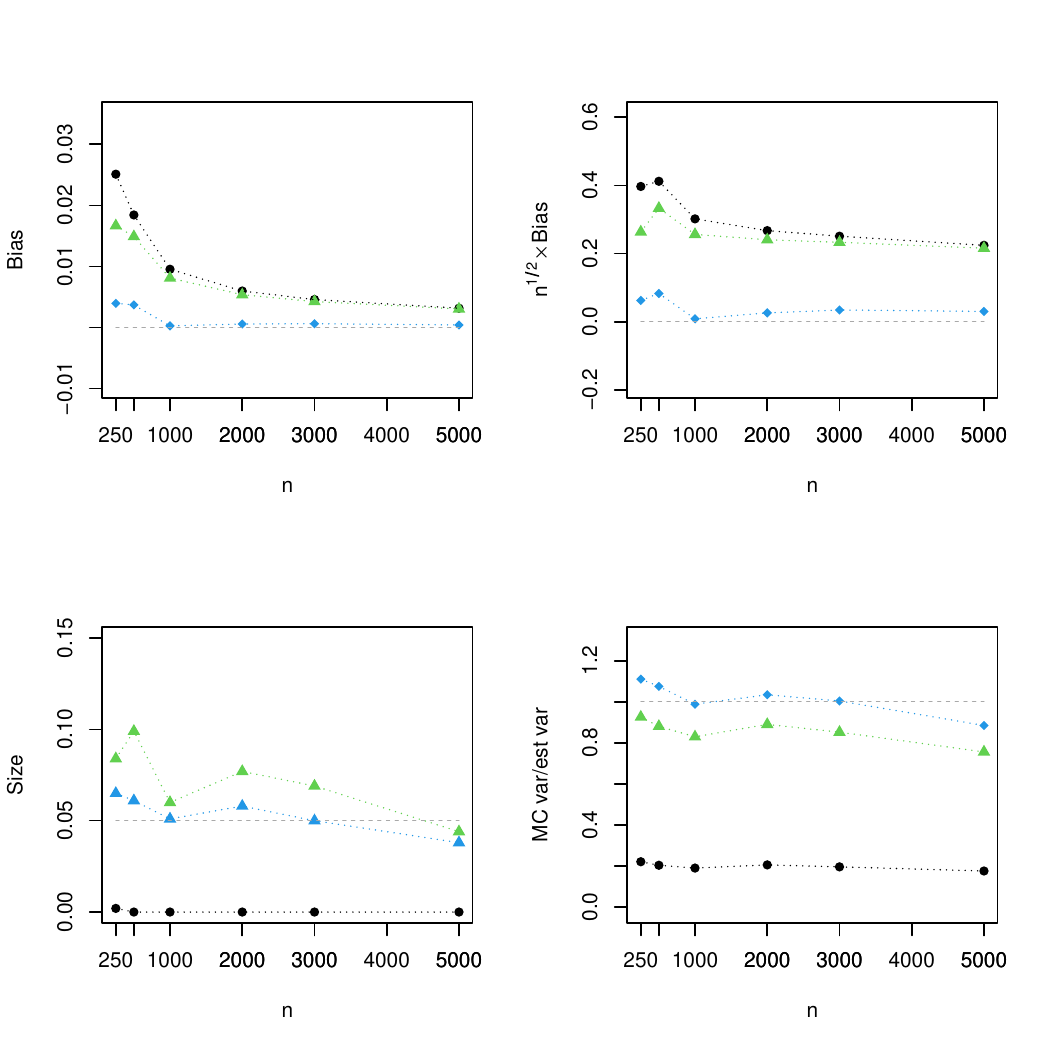}
\end{figure}

\begin{figure}
  \caption{Second experiment (outcome model estimated correctly). \textcolor{red}{Red} squares represent `OR'; \textcolor{green}{green} triangles represent `DML'; \textcolor{blue}{blue} triangles with represent `DR-DML'.}
  \centering
    \includegraphics[width=0.55\textwidth]{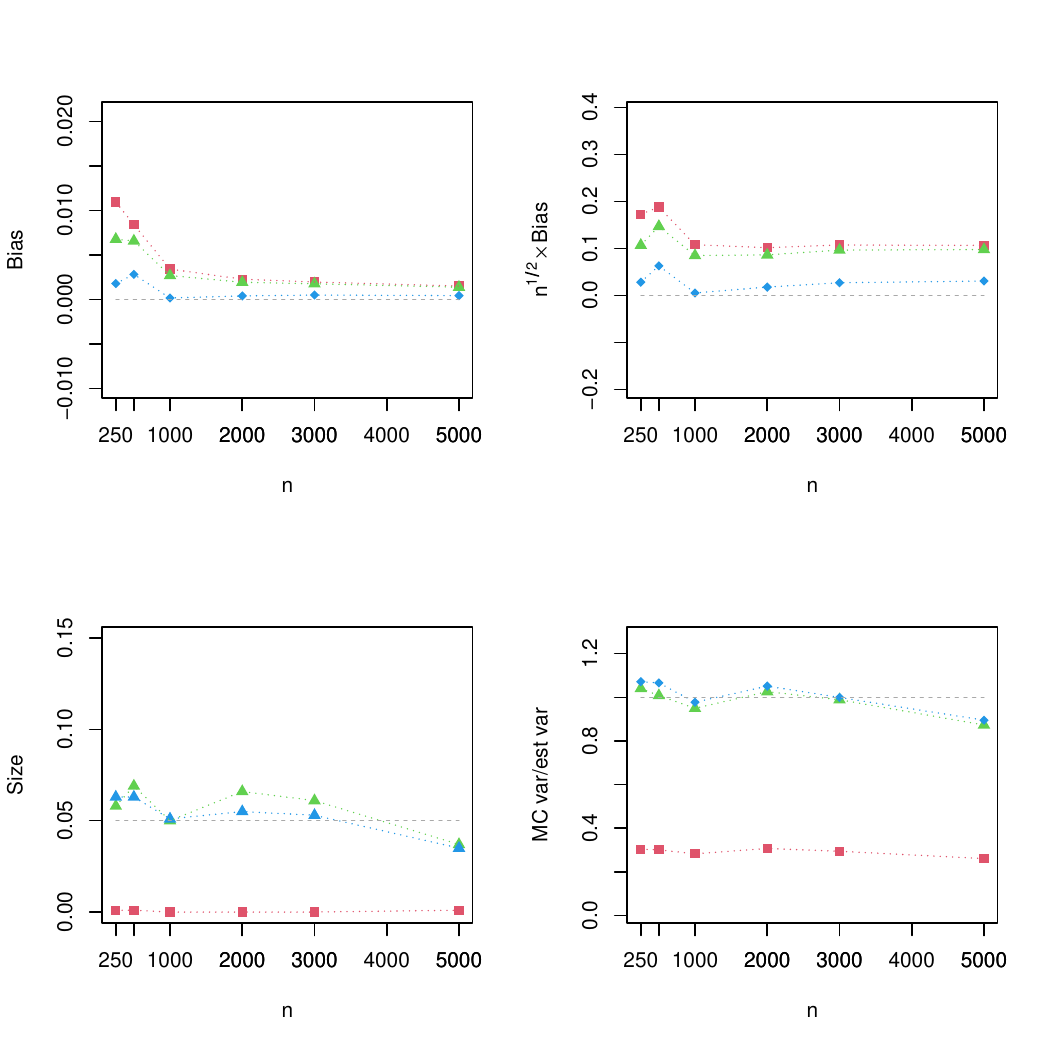}
\end{figure}

\newpage
\bibliography{fRDMLbibfile}

\end{document}